\documentclass[pra,twocolumn,showpacs,preprintnumbers,amsmath,amssymb]{revtex4-1}
\usepackage{graphicx}

\begin{document}
\title{ Quantum magnetism of spinor bosons in optical lattices with synthetic non-Abelian gauge fields  }
\author{ Fadi Sun$^{1,2}$, Jinwu Ye,$^{2,3}$ and Wu-Ming Liu$^{1}$  }
\affiliation{
$^{1}$Beijing National Laboratory for Condensed Matter Physics,
Institute of Physics, Chinese Academy of Sciences,
Beijing 100190, China   \\
$^{2}$Department of Physics and Astronomy, Mississippi State University, MS, 39762, USA  \\
$^{3}$Key Laboratory of Terahertz Optoelectronics, Ministry of Education, Department of Physics, Capital Normal University, Beijing, 100048, China  }

\date{\today }

\begin{abstract}
 We study quantum magnetism of interacting spinor bosons at integer fillings hopping in a square lattice in the presence of
 of non-Abelian gauge fields. In the strong coupling limit, it leads to the Rotated ferromagnetic Heisenberg  model (RFHM) which
 is a new class of quantum spin model.
 We introduce Wilson loops to characterize frustrations and gauge equivalent classes.
 For a special equivalent class, we identify a new spin-orbital entangled commensurate ground state.
 It supports not only commensurate magnons, but also a new gapped elementary excitation: in-commensurate magnons with two gap minima continuously
 tuned by the SOC strength. At low temperatures, these magnons lead to dramatic effects in many physical quantities such as
 density of states, specific heat, magnetization, uniform susceptibility, staggered susceptibility and various spin correlation functions.
 The commensurate magnons lead to a pinned central peak in the angle resolved light or atom Bragg spectroscopy. However,
 the in-commensurate magnons  split it  into two located at their two gap minima.
 At high temperatures, the transverse spin structure factors depend on the SOC strength explicitly.
 The whole set of Wilson loops can be mapped out by measuring the specific heat at the corresponding orders
 in the high temperature expansion.
 We argue that one gauge may be realized in current experiments and other gauges may also be realized in near future experiments.
 The results achieved along the exact solvable line sets up the stage to investigate dramatic effects when tuning away from it by various means.
 We sketch the crucial roles to be played by these magnons
 at other equivalent classes, with spin anisotropic interactions and in the presence of finite magnetic fields.
 Various experimental detections of these new phenomena are discussed. Rotated Anti-ferromagnetic Heisenberg  model are also briefly mentioned.
\end{abstract}


\maketitle

\section{Introduction}
Quantum magnetism has been an important and  vigorous research field in material science for many decades \cite{scaling,aue}.
In general, Heisenberg model and its variants have been widely used to study quantum magnetisms in both kinds of systems.
However, they can not be used to describe materials or cold atom systems with strong spin-orbit couplings (SOC).
Recently the investigation and control of spin-orbit coupling (SOC) have become subjects of
intensive research in both condensed matter and cold atom systems after the discovery of the topological insulators \cite{kane,zhang}.
In the condensed matter side, there are  increasing number of new quantum materials with significant SOC,
including several new 5d transition metal oxides and heterostructures of transition metal systems \cite{kitpconf}.
In the cold atom side,  there have also been impressive advances in
generating artificial gauge fields in both continuum  and on optical lattices \cite{sorev}.
Several experimental groups have successfully generated
     a 1D synthetic non-Abelian gauge potential coupled to neutral atoms by dressing internal atomic spin states with spatially varying Laser beams \cite{sorev}.
     Unfortunately, so far,  2D  Rashba or Dresselhauss SOC  and 3D isotropic (Weyl) SOC  have not been implemented experimentally.

     Notably, there are very recent remarkable advances to generate magnetic fields in optical lattices\cite{stagg1,stagg2,uniform1,uniform2,uniform3,newexp,newexp2,sorev}.
     Indeed, staggered magnetic field along one direction (Fig.7a) \cite{sorev} in an optical lattice has been achieved by using Laser assisted tunneling in superlattice potentials \cite{stagg1} and by dynamic lattice shaking \cite{stagg2}.
 By using laser-assisted tunneling in a tilted optical lattice through periodic driving with a pair of far-detuned running-wave beams,
 One experimental group \cite{uniform1} (see also \cite{uniform2,uniform3} for related work)
 successfully generated the time-reversal symmetric Hamiltonian underlying the quantum spin Hall effects (Fig.7b):
 namely, two different pseudo-spin components
 (two suitably chosen hyperfine states for $^{87}$Rb  atoms) experience opposite directions of the uniform magnetic field.
 In one recent experiments \cite{newexp}, both the vortex phase and Meissner phase were observed for weakly interacting bosons
 in the presence of strong artificial magnetic field in an optical lattice ladder systems.
 In another \cite{newexp2}, a first measurement on Chern number of bosonic Hofstadter bands was also performed.
 The celebrated Haldane model was also realized for the first time with ultracold fermions \cite{haldane}.
 As pointed out in \cite{uniform3},  Non-Abelian gauge in Eq.\eqref{bosonint} can be achieved by adding spin-flip Raman lasers to
 induce a $ \alpha \sigma_x $ term along the
 horizontal bond, or by driving the spin-flip transition with RF or microwave fields.
 Scaling functions for both gauge-invariant and non-gauge invariant quantities across topological transitions of non-interacting fermions
 driven by the non-Abelian gauge potentials on an optical lattice
 have also been derived \cite{tqpt}. However, so far,  possible new class of quantum magnetic phenomena due to the interplay
 among the interactions, the SOC and lattice geometries have not been addressed yet.


In this paper, we investigate such an interplay systematically 
by studying the system of interacting spinor (multi-component) bosons 
at integer fillings hopping in a square lattice in the presence of SOC.
Starting from spinor boson Hubbard model in the presence of non-Abelian gauge fields,
at strong coupling limit, we derive a Rotated Ferromagnetic Heisenberg model (RFHM)
which is a new class of quantum spin models to describe 
cold atom systems or materials with strong SOC.
Wilson loops are introduced to characterize frustrations 
and gauge equivalent classes in this RFHM.
For a special equivalent class, we enumerate all the discrete symmetries,
especially discover a hidden spin-orbit coupled continuous U$(1)$ symmetry,
then we identify a new commensurate spin-orbital entangled quantum ground state 
and classify its symmetry breaking patterns.
By performing spin wave expansion (SWE) above the ground state, 
we find that it supports two kinds of gapped excitations as the SOC parameter changes:
one is commensurate magnons C-C$_0$, C-C$_{\pi}$ 
with one gap minimum pinned at $(0,0)$ or $(0,\pi)$,
another is a novel elementary excitation: in-commensurate magnons C-IC
with two gap minima $(0,\pm k^{0}_y)$ continuously tuned by the SOC strength.
The boundary between the two kinds of magnons are signaled by its divergent effective mass
(or equivalently divergent density of states (DOS)).
Both kinds of magnons lead to dramatic experimental observable consequences 
in many thermodynamic quantities such as the magnetization, specific heat, uniform and staggered susceptibilities, the Wilson ratio and also spin correlation functions such as the uniform and staggered, dynamic and equal-time, longitudinal and transverse, normal and anomalous spin-spin correlation functions.
At low temperatures, we determine the leading temperature dependencies in the 
C-C$_0$, C-$C_{\pi}$ regime and C-IC regime, also near their boundaries.
The magnetization leads to one sharp peak in the {\sl longitudinal} 
equal-time spin structure factors at $ (\pi,0) $.
Both kind of magnons lead to sharp peaks in {\sl dynamic transverse } spin correlation functions.
The commensurate magnons lead to one Gaussian peak in the {\sl transverse} equal-time spin structure factors with its center pinned at $(0,0)$  or $(0,\pi)$ respectively.
However, the in-commensurate magnons splits the peak into two centered at their two gap minima $ (0,\pm k^{0}_{y}) $ continuously tuned by  the SOC strength.
At high temperatures, by performing high temperature expansion, we find that the equal-time transverse spin structure factors depend on SOC strength explicitly,
the specific heat depends on all sets of Wilson loops at corresponding orders in the high temperature expansion.
This fact sets up the principle to map out the whole sets of Wilson loops by specific heat measurements.
Experimental detections by atom or light Bragg spectroscopies \cite{lightatom1,lightatom2} and specific heat measurements are discussed.
We argue that a special gauge (called ``U(1)'' gauge) may be achieved by a combination of previous experiments to realize staggered magnetic field \cite{stagg1,stagg2} and recent experiments to realize quantum spin Hall effects \cite{uniform1,uniform2,uniform3}.
It is also possible to realize the other gauges in near future experiments.

The results achieved on the special equivalent class sets up the stage to investigate dramatic effects when tuning away from it by adding or changing
various parameters.
Especially, the crucial roles played by these magnons in the RH model at generic equivalent classes,
or with spin anisotropic interactions or in the presence of finite uniform and staggered magnetic fields
will also be briefly mentioned.

The paper was organized as follow.
In Sec.II, starting from the spinor boson Hubbard model in the presence of non-Abelian gauge fields (Fig.1a),
in the strong coupling limit, we derive the Rotated Ferromagnetic Heisenberg model (RFHM),
also stress its crucial differences than the previously well known modes such as
Heisenberg model \cite{scaling,aue}, Kitaev model \cite{kit,hk3}, Dzyaloshinskii-Moriya (DM) interaction \cite{dmterm1,dmterm2}  and some other strong coupling models \cite{strong1,strong2}.
In Sec.III, we introduce the Wilson loops (Fig.1b) to characterize gauge equivalent classes and frustrations of the RFHM.
We identify an exactly solvable line in the non-Abelian gauge parameter space
and also determine all the discrete symmetries, especially a hidden spin-orbital coupled continuous U$(1)$ symmetry.
We determine the exact ground state (Fig.2a) and its symmetry breaking patterns.
In Sec.IV, by using spin wave expansion (SWE), we will determine
the excitation spectra of commensurate magnons and in-commensurate magnons (Fig.2b,4).
We will also compute their contributions to the many thermodynamic quantities such as
the magnetization, specific heat, uniform and staggered susceptibilities, the Wilson ratio
and also the finite temperature phase diagrams (Fig.3).
In Sec.V, we determine all the spin-correlation functions such as
the uniform and staggered, dynamic and equal-time, longitudinal and transverse, normal
and anomalous spin-spin correlation functions.
We use the hidden spin-orbital coupled continuous U$(1)$ symmetry
to derive exact relations among different spin correlation functions.
We specify how the In-commensurate magnons
will split the equal-time spin structure factors into two peaks located at their two gap minima (Fig.5).
We stress the asymmetric shape of the uniform normal spin structure factor
which can be measured by light or atom scattering cross-section.
In Sec.VI, using high temperature expansion, we will evaluate specific heat and equal-time spin structure factors.
We stress that in principle, the whole set of Wilson loops can be measured
by the specific heat measurements at high temperatures.
In Sec. VII, we perform a local gauge transformation to a basis
where the hidden spin-orbital U$(1)$ symmetry becomes an explicit U$(1)$ symmetry.
We contrast the gauge field configurations in the U$(1)$ basis (Fig.7a)
against Quantum spin Hall effects (Fig.7b) realized in recent experiments \cite{uniform1,uniform2,uniform3}.
We propose a scheme how the U$(1)$ basis can be achieved by some possible combinations of previous experiments
to realize staggered magnetic fields \cite{stagg1,stagg2} and recent  experiments to realize Quantum spin Hall effects.
All the thermodynamic quantities are gauge invariant
(up to some exchange between uniform and staggered susceptibilities),
but spin correlation functions are not.
In Sec.VIII, we re-evaluate all the spin correlation functions in the U$(1)$ basis at both low and high temperatures,
then contrast with those in the original basis.
In addition to its potential to be more easily realized in near future experiments,
another advantage of the U$(1)$ basis is that
the asymmetry in the light or atom scattering cross sections in the original basis (Fig.5)
can be eliminated in the U$(1)$ basis (Fig.8 and 9),
so all the  commensurate magnons and in-commensurate magnons can be more easily detected in the U$(1)$ basis.
In the conclusion Sec.IX, we discuss experimental realizations of the RFHM, higher order corrections in the SWE and
a possible in-commensurate superfluid at weak coupling $ U/t \ll 1 $. We also
stress the important roles of these magnons in driving quantum phase transitions when tuning away from the solvable line by various means
changing $(\alpha,\beta)$, spin anisotropic interactions and external magnetic fields.
Some technical details are presented in the four appendixes.
{\sl All the physical quantities shown in all the figures are made dimensionless. }

\section{Synthetic Rotated spin-$S$ Heisenberg model in the strong coupling limit}

The pseudo-spin 1/2 boson Hubbard model at integer fillings
$ \langle b^{\dagger}_{\uparrow} b_{\uparrow}+ b^{\dagger}_{\downarrow} b_{\downarrow} \rangle =N $
subject to a non-Abelian gauge potential is \cite{tqpt}:
\begin{equation}
	H_{b} =  -t\sum\limits_{\langle i,j\rangle}
	[b^\dagger(i\sigma)
	U_{ij}^{\sigma\sigma'}
	b(j\sigma')+h.c.]
	+\frac{U}{2}\sum_{i}(n_{i}-N)^2
\label{bosonint}
\end{equation}
where $ \sigma = \uparrow, \downarrow $ stands for the two hyperfine states which   are $ | F,m_F \rangle=|1,-1 \rangle, |2,-1 \rangle $  used in  \cite{uniform1} or  $ | 2,2 \rangle, |2, -2 \rangle $  used in \cite{uniform3},
the $U_1=e^{i \alpha \sigma_x}$, $U_2=e^{i \beta \sigma_y}$
are the non-Abelian gauge fields put on the two links in the square lattice (Fig.1a),
$n_i=n_{i \uparrow}+n_{i \downarrow}$ is the total density.
In this paper, we focus on spin-independent interaction.
This is probably the most relevant experimental situation,
because the spin-dependent energies are typically much smaller than the on-site interaction.
However, the dramatic effects of spin-dependent interactions Eq.\eqref{anisotropy} will be mentioned in the Sec.VII and the conclusion section.
Following \cite{tqpt}, we find the Wilson loop around one square
$ W_b={\rm Tr}[ U_1 U_2 U^{-1}_{1} U^{-1}_{2}]=2-4 \sin^{2}\alpha \sin^{2} \beta $.
The $ W_b= \pm 2 $ ($ |W| < 2 $) correspond to Abelian $ \theta=0,\pi $ (non-Abelian) regimes (Fig.1b).
Similar to \cite{tqpt}, the other two Wilson loops around two squares oriented along $ x $ and $ y $ axis
are $ W_{b,x}=2-4 \sin^{2} 2\alpha \sin^{2} \beta$, $W_{b,y}= 2-4 \sin^{2} \alpha \sin^{2} 2\beta $.
In the following, we focus on the strong coupling limit $ U \gg t $.
The possible superfluid states at weak $U\ll t$ coupling will be briefly mentioned in the conclusion section.

\begin{figure}
\includegraphics[width=8.5cm]{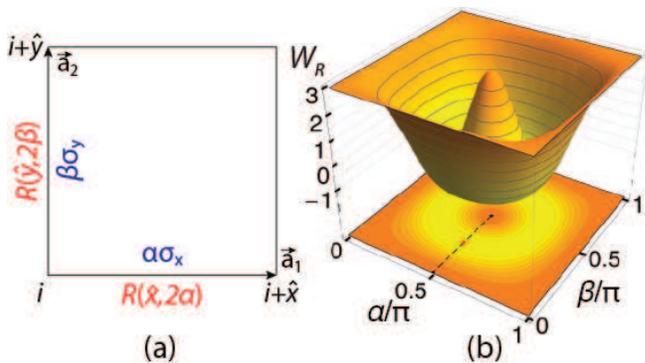}
\caption{(Color online)
(a) For the bosonic model Eq.\eqref{bosonint} 
[the Rotated Heisenberg (RH) quantum spin model Eq.\eqref{rh}],
the non-Abelian gauge potentials $U_{1}=e^{i \alpha \sigma_x}$, 
$U_{2}=e^{i \beta \sigma_y} $ (blue or dark gray)
[the two rotation matrices $ R_x, R_y $ (red or light gray)]
with directions are put on the two links $x$, $y$ inside the unit cell respectively.
(b) Wilson loop $ W_R(\alpha,\beta) $ of the RH model Eq.\eqref{rh} 
reaches maximum ,$3$, at the Abelian points,
minimum ,$-1$, in the most frustrated regime.
Shown at the bottom is the dashed line $(\alpha=\pi/2,\beta)$ focused in this paper.
The $\times$ stands for the most frustrated point $ \beta=\pi/4 $.
 }
\label{fig1}
\end{figure}

In the strong coupling limit $ U \gg t $, to leading order in $ t^2/U $,
we get a spin $S=N/2$ ``rotated''  Ferromagnetic Heisenberg (RFH) model:
\begin{equation}
	H_{RH} \!=\! -J\!\sum_{i}[S^{a}_{i}R^{ab}\!(\hat{x},\!2\alpha) S^{b}_{i+\hat{x}}
               		      +S^{a}_{i}R^{ab}\!(\hat{y},\!2\beta)  S^{b}_{i+\hat{y}}]
\label{rh}
\end{equation}
with a ferromagnetic (FM) interaction $ J= 4 t^2/U  $
and the sum is over the unit cell $i$ in Fig.1a,
the $ R(\hat{x}, 2 \alpha)$,  $R(\hat{y}, 2 \beta)$ are
two SO$(3)$ rotation matrices around  
the $ \hat{x}, \hat{y} $  spin axis by angle $ 2\alpha$, $2\beta $
putting on the two  bonds  $x$,$y$ respectively (Fig.1a).
Obviously,  at $ \alpha=\beta =0 $, the Hamiltonian becomes the usual FM Heisenberg model
$ H= -J \sum_{ \langle i j \rangle } \mathbf{S}_{i} \cdot  \mathbf{S}_{j} $.
In fact, when expanding the two $R$ matrices, one can see that Eq.\eqref{rh} leads to
a Heisenberg \cite{scaling} + Kitaev  \cite{kit,hk3} + DM interaction \cite{dmterm1,dmterm2}:
$ H_{s}= - J [\sum_{\langle i j \rangle  } J^a_{H} \mathbf{S}_{i} \cdot \mathbf{S}_{j}
	+\sum_{\langle i j \rangle a } J^{a}_{K} S^{a}_{i} S^{a}_{j}
	+\sum_{\langle i j \rangle a } J^{a}_{D} \hat{a} \cdot \mathbf{S}_{i} \times \mathbf{S}_{j} ] $
where $ \hat{a}= \hat{x}, \hat{y} $, $ J^{x}_H=\cos 2 \alpha,  J^{y}_H=\cos 2 \beta $;
$ J^{x}_K= 2 \sin^2 \alpha,  J^{y}_K= 2 \sin^2 \beta $ and $ J^{x}_D=\sin 2 \alpha,  J^{y}_D=\sin 2 \beta $.
However, as we show in the following,
many deep physical pictures and exact relations can only be established
in the $R$-matrix representation Eq.\eqref{rh}.

Note that there are other strong coupling models.
For example, Ref. \cite{strong1} studied the effects of $U$ on Kane-Mele model \cite{kane,zhang}
(called Kane-Mele-Hubbard model with the $S^z$ conserving SOC),
focusing on the stability of topological insulator
and the corresponding helical edge states against the interactions $ U $.
Ref.\cite{strong2} studied time-reversal invariant Hofstadter-Hubbard model of
spin $1/2 $ fermions hopping on a square lattice subject to an Abelian flux $\alpha=p/q$.
This is the quantum spin Hall effects model in Fig.7b.
The RH model Eq.\eqref{rh} is in a completely different class than these models.
It will be contrasted with the quantum Spin Hall effects in Sec. VIII.
The Rotated anti-ferromagnetic Heisenberg (RAF) model with $J= -4t^2/U $
will be mentioned in the conclusion section.

\section{ Classification by Wilson loops and an exactly solvable line }

The advantages of RHM form in Eq.\eqref{rh} is significant: 
It is much more than its beauty and elegancy,
it contains deep and important physics and lead to many important physical consequences.
Only in this representation, one can introduce Wilson loops $W_R$ for the quantum spin models
to characterize equivalent classes in the quantum spin models. 
The Wilson loops $W_R$ can be used to establish many highly non-trivial exact relations presented in the whole paper and also  the four appendixes.
These exact relations are extremely important to put various constraints 
on any practical calculations such as spin wave expansion (SWE) in the next section.

The $R$-matrix Wilson loop $ W_R $ around a fundamental square (Fig.1a) is defined as 
$W_{R}={\rm Tr}[R_x R_y R^{-1}_x R^{-1}_y ]=
    [\cos(2\alpha)+\cos(2\beta)-\cos(2\alpha)\cos(2\beta)]
	[2+\cos(2\alpha)+\cos(2\beta)-\cos(2\alpha)\cos(2\beta)] $
to characterize the equivalent class and frustrations in the RH model Eq.\eqref{rh}.
The $ W_R=3 $ ($W_R \neq 3 $) stands for the Abelian (non-Abelian) points (Fig.1b).
For example, all the 4 edges and the center belong to Abelian points $ W_R=3 $.
All the other points belong to Non-Abelian points (Fig.1b).
The other two Wilson loops  around two squares oriented along $x$ and $y$ axis are
$ W_{Rx}=[\cos(4\alpha)+\cos(2\beta)-\cos(4\alpha)\cos(2\beta)]
	[2+\cos(4\alpha)+\cos(2\beta)-\cos(4\alpha)\cos(2\beta)] $
and
$ W_{Ry}=[\cos(2\alpha)+\cos(4\beta)-\cos(2\alpha)\cos(4\beta)]
	[2+\cos(2\alpha)+\cos(4\beta)-\cos(2\alpha)\cos(4\beta)] $.
The relations between two sets of Wilson loops in Eq.\eqref{bosonint}  and \eqref{rh}
are two to one relation due to the coset SU$(2)$/Z$_2=$ SO$(3)$.
For example, the Abelian points $ W = \pm2 $ correspond to  $ W_R=3 $.
We stress that
{\sl any RH model with the same set of Wilson loops can be transformed to each other
by performing local SO$(3)$ transformations}
and belong to the same equivalent class.
As shown in the following, the classification according to the Wilson loops
can be used to establish connections among seemly {\sl different } phases.
Most importantly, as shown in Sec.VI, we show that the whole set of Wilson loops can be mapped out by measuring the specific heat
at the corresponding orders in the high temperature expansion.

In the $S\rightarrow\infty$ limit, the RH model Eq.\eqref{rh} becomes classical.
Some interesting results on the possible rich classical ground states
at some sets of general $(\alpha,\beta)$ in the Heisenberg-Kitaev-DM representation
were attempted numerically in \cite{classdm1,classdm2}.
Here, we plan to study the quantum phenomena in the RH model at generic $(\alpha,\beta)$.
However, it is a very difficult task, so we take a ``divide and conquer'' strategy.
First, we identify an exact solvable line:
the dashed line $\alpha=\pi/2$,$0<\beta<\pi/2$ in Fig.1b
and explore new and rich quantum phenomena along the line.
Then starting from the deep knowledge along the solvable line,
we will try to investigate the quantum phenomena at generic $(\alpha,\beta)$.
In this paper, we will focus on the first task.
The second task will be briefly mentioned in the conclusion section (Fig.10) and presented in details elsewhere.
In the past, this kind of ``divide and conquer'' approach has been very successful
in solving many quantum spin models.
For example, in single (multi-)channel Kondo model, one solve the Thouless  (Emery-Kivelson) line \cite{kondo1,kondo2}, then do perturbation away from it.
In quantum-dimer model, one solves the Rohksa-Kivelson (RK) point which shows spin liquid physics \cite{dimer}, then one can study the effects of various
perturbations away from it \cite{dimer2}.
For the Heisenberg-Kitave (HK) model \cite{kitpconf} and its various extensions,
one solves the FM or AFM Kitaev point \cite{kit,hk3} which shows spin liquid and non-Abelian statistics.
Then on can study various Kitaev materials away from the Kitaev point.

The Wilson loops along the dashed line are
$W= 2 \cos 2 \beta \neq \pm 2$, $W_x=2$, $W_y=2-4 \sin^2 2 \beta $ in $ H_b $ and
$W_R=2 \cos 4\beta + 1 \neq 3$, $W_{R_x}=3$, $W_{R_y}=4 \cos^2 4 \beta-1 $ in $ H_{RH} $.
So all the points along the dashed line except at the two Abelian points $ \beta=0,\pi/2 $
display dramatic non-Abelian effects.
At the two ends of the dashed line $ \alpha=\pi/2$, $\beta=0$ ($\beta=\pi/2 $) in Fig.1b,
we get the FM Heisenberg model in the rotated basis
$ H= -J \sum_{\langle ij\rangle } \tilde{\mathbf{S}}_{i} \cdot \tilde{\mathbf{S}}_{j} $,
where the  $\tilde{\mathbf{S}}_{i} = R(\hat{x},\pi n_1) \mathbf{S}_{i}$
($ \tilde{\mathbf{S}}_{i} = R(\hat{x},\pi n_1)  R(\hat{y},\pi n_2) \mathbf{S}_{i} $).
One can also see $ W_R(\beta)=W_R( \pi/2-\beta) $
which indicates $ \beta $ and $ \pi/2- \beta $ can be related by some local rotations.
Indeed, it can be shown that under the local rotation
$ \tilde{\mathbf{S}}_{i} = R(\hat{x},\pi ) R(\hat{y},\pi n_2) \mathbf{S}_{i}$,
$ \beta \rightarrow \pi/2 - \beta $.
The most frustrated point with $ W_R=-1 $ is  located at the middle point $ \beta=\pi/4 $ (Fig.1b).
One can also show that $ \sum_i(-1)^{i_x} S^{y}_{i} $ is a conserved quantity
$[H_b,\sum_i(-1)^{i_x}b_i^\dagger\sigma^y b_i]=0 $.
This spin-orbit coupled U$(1)$ symmetry  will become transparent
after a local gauge transformation to the ``U$(1)$'' basis in Eq.\eqref{kinetic1}.
Obviously, this spin-orbit coupled U$(1)$ symmetry is kept in the RH model Eq.\eqref{rh}
$[H_{RH},\sum_i(-1)^{i_x}S^{y}_{i}]=0$.
It will be used to identify the exact {\sl quantum} ground state
and also establish exact relations among various spin correlations functions.

It is convenient to make a $R_x(\pi/2)$ rotation to rotate spin $Y$  axis to $Z$ axis
(More directly, one can just put $ \beta \sigma_z $  along the $y$ bonds in Fig.1a.),
then the Hamiltonian Eq.\eqref{rh} along the dashed line can be written as
\begin{align}
	H_{d} & =  -J\sum_{i}[\frac{1}{2}(S_i^+S_{i+x}^++S_i^-S_{i+x}^-)-S_i^zS_{i+x}^z
\nonumber   \\
	  & + \frac{1}{2}(e^{i2\beta}S_i^+S_{i+y}^-+e^{-i2\beta}S_i^-S_{i+y}^+)+S_i^zS_{i+y}^z]
\label{originalH}
\end{align}
All the possible symmetries of $H_{d}$ are analyzed in the appendix A.
It is shown in the appendix B that the $Y$-$x$ state with the ordering wave vector $(\pi,0)$ 
(Fig.2a)
is the exact ground state  with the ground state energy $ E_0= -2 N J S^2 $.
The conserved quantity $  \sum_i(-1)^{i_x} S^{y}_{i} $ reaches its maximum value $NS$ in the ground state.
The symmetry breaking patterns of the $Y$-$x$ state is analyzed in appendix B.

\begin{figure}[htb]
\includegraphics[width=8.5cm]{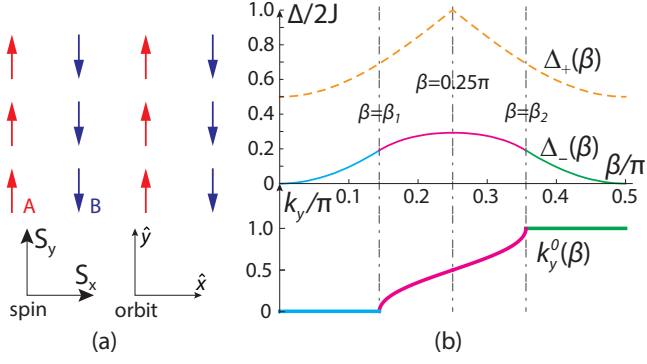}
\caption{(Color online)
(a) The exact ground state is the $Y$-$x$ state
where the first capital letter indicates spin polarization along $Y$ direction,
the second small letter indicates the orbital ordering along the $x$ bond.
(b) The minima position $\mathbf{k}_0=(0, \pm k^{0}_y)$ in the RBZ of the acoustic branch
and its gap $ \Delta_{-}(\beta) $ at the minima.
When $ 0\leq \beta< \beta_1 =\arccos[\sqrt{1+\sqrt{5}}/2]\approx 0.144\pi $,
there is one minimum pinned at $ k^{0}_y=0 $ with the gap $\Delta_{-}(\beta)=\sin^2\beta $.
When $ \beta_1\leq\beta<\beta_2=\pi/2-\beta_1 $,
there are two minima at
$  \pm k^{0}_y= \pm \arccos[{\sqrt{1+\sin^22\beta}}/{\tan2\beta}] $
with the gap $  \Delta_{-}(\beta)=1-{\sqrt{1+\sin^22\beta}}/{(2\sin2\beta)} $.
Only $  k^{0}_y > 0 $ is shown here.
When $ \beta_2\leq\beta<\pi/2 $,
there is one minimum at $k^{0}_y= \pm \pi$ with the gap $\Delta_{-}(\beta)=\cos^2\beta $.
The $ \Delta_{+}(\beta) $ is the minima gap of the optical branch.
When $\beta<\pi/4$, the minimum is $ k^{u}_0=(\pi/2,0)$
with the gap $ \Delta_{+}(\beta)=1-\frac{1}{2}\cos2\beta$.
When $\beta>\pi/4$, the minimum is $k^{u}_0=(\pi/2,\pi)$
with $ \Delta_{+}(\beta) =1+\frac{1}{2}\cos2\beta$.
The gaps of both branches reach maximum at the most frustrated point $ \beta=\pi/4 $ in Fig.1b. }
\label{zx}
\end{figure}

\section{ Thermodynamic quantities at low temperatures }

In this section, by using spin wave expansion (SWE)\cite{sw1,sw12-1,sw12-2,sw2,sw3},
we will first discover C-C$_0$, C-C$_{\pi}$ and C-IC magnons,
then evaluate their contributions to the Magnetization,
uniform and staggered susceptibilities,
specific heat and Wilson ratio at low temperatures

\subsection{ Commensurate and In-Commensurate  magnons }

Introducing the Holstein-Primakoff (HP) bosons \cite{sw1,sw12-1,sw12-2,sw2,sw3}
$ S^{+}=  \sqrt{2S-a^{\dagger} a }a$,
$S^{-}= a^{\dagger} \sqrt{2S-a^{\dagger} a }$,
$S^{z}=S- a^{\dagger} a $
for sublattice A and
$ S^{+}= b^{\dagger} \sqrt{2S-b^{\dagger} b }$,
$S^{-}=  \sqrt{2S-b^{\dagger} b }b$,
$S^{z}=b^{\dagger} b-S $
for the sublattice B in Fig.2a.
By a unitary transformation in $ \mathbf{k} $ space:
\begin{equation}
	\begin{pmatrix}
		a_\mathbf{k}\\
		b_\mathbf{k}\\
	\end{pmatrix}
	=
	\begin{pmatrix}
		\sin\frac{\theta_\mathbf{k}}{2} &\cos\frac{\theta_\mathbf{k}}{2}\\
		-\cos\frac{\theta_\mathbf{k}}{2} &\sin\frac{\theta_\mathbf{k}}{2}\\
	\end{pmatrix}
	\begin{pmatrix}
		\alpha_\mathbf{k}\\
		\beta_\mathbf{k}\\
	\end{pmatrix}
\label{unitary}
\end{equation}
where $
	\sin\theta_\mathbf{k}
	=\frac{\cos k_x}{\sqrt{\cos^2k_x+\sin^22\beta\sin^2k_y}},
	\quad
	\cos\theta_\mathbf{k}
	=\frac{\sin2\beta\sin k_y}{\sqrt{\cos^2k_x+\sin^22\beta\sin^2k_y}}
$, the Hamiltonian $ H_{d} $ can be diagonalized:
\begin{equation}
	H_{m}  = E_0 + 4JS\sum_\mathbf{k} 
	[E_+(\mathbf{k})\alpha_\mathbf{k}^\dagger\alpha_\mathbf{k}
	+E_-(\mathbf{k})\beta_\mathbf{k}^\dagger\beta_\mathbf{k}]
\label{energypm}
\end{equation}
where $ E_0= - 2 NJ S^2 $ and $ \mathbf{k} $ belongs to the reduced Brillioun zone (RBZ)
and $ E_{\pm}(k) = 1-\frac{1}{2}\cos2\beta\cos k_y
\pm\frac{1}{2}\sqrt{\cos^2k_x+\sin^2 2\beta\sin^2 k_y} $
are the excitation spectra of the acoustic and optical branches respectively.
Note that $ \sin\theta_k $ is even under the space inversion 
$ \mathbf{k} \rightarrow -\mathbf{k} $,
but $ \cos\theta_\mathbf{k} $ is odd.

At the two Abelian points $ \beta=0, \pi/2 $, as shown above,
the system has SU$(2)$ symmetry in the correspondingly rotated basis,
Eq.\eqref{energypm} reduces to the FM spin wave excitation spectrum $ \omega \sim k^2 $
at the minimum $ (0,0) $ and $ (0,\pi) $ respectively.
The positions of the minima and the gap at the minima of both branches are shown in Fig.2b.
One can see that the $Y$-$x$ ground state supports two kinds of gapped excitations.
(1) When $ 0 < \beta < \beta_1 $,
it supports commensurate magnons C-C$_0$ with one gap minimum pinned at $ (0,0) $.
Here, we use the first letter to indicate the ground state, the second the excitations.
Similarly, when $ \beta_2 < \beta < \pi/2 $,
commensurate magnons C-C$_{\pi}$ with one gap minimum pinned at $ (0, \pm \pi ) $.
(2) In the middle regimes $ \beta_1<\beta < \beta_2 $,
it supports in-commensurate magnons C-IC  
with two continuously changing gap minima at $ (0,\pm k^{0}_{y} ) $
tuned by the SOC strength (Fig.3).
In fact, at the most frustrated point $ \beta=\pi/4 $,
there are two gap minima $ \pm k^{0}_{y}=\pm \pi/2 $
which indicates a $ 2 \times 4 $ short-ranged commensurate orbital structure,
but there is no pinned plateau near this point.
In general, $ k^{0}_{y} $ is an irrational number at $ \beta_1 < \beta < \beta_2 $,
so justify the name C-IC .
Both kinds of magnons have striking experimental consequences
in all the thermodynamic quantities at finite $T$ to be discussed in the following.

\subsection{ Magnetization, specific heat, uniform and staggered susceptibilities  and Wilson ratio. }

\begin{figure}[htb]
\includegraphics[width=7.5cm]{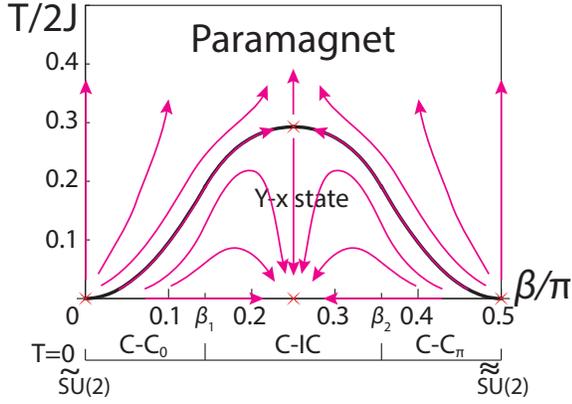}
\caption{(Color online)
The finite temperature  phase diagram along the dashed line in Fig.1b.
Along the dashed line, the $Y$-$x$ ground state supports
C-C$_0$, C-IC, C-C$_{\pi}$ magnons consecutively.
There is an enlarged symmetry at $\beta=\pi/4$.
The finite temperature phase transitions are controlled by the renormalization
group (RG) flow fixed point at $(\beta=\pi/4,T_m)$  
where $ T_m $ is the maximum temperature at $ \beta=\pi/4$.
Its universality class will be speculated in Sec. VIII-C. 
The arrows indicate the RG flows.
}
\label{finiteT}
\end{figure}

At the two Abelian points, at any finite $T$,
the spin wave fluctuations will destroy the FM order as dictated
by the Mermin-Wegner theorem (Fig.3).
However, at any non-Abelian points along the dashed line,
although the ground state remains the $Y-x$ ground state (Fig.2a),
there is a gap $\Delta_{-}(\beta)$ in the excitation spectrum,
so the order survives up to a finite critical temperature
$ T_c \sim  \Delta_{-}(\beta) $ (Fig.3).
At low temperatures $ T < T_c $ in Fig.3, one can ignore the optical branch.
Expect at $ \beta_1 ( \beta_2=\pi/2-\beta_1 ) $, the acoustic branch can be expanded around the minima
$ \mathbf{k}= \mathbf{k}_0+\mathbf{q} $
as $E_-(\mathbf{q};\beta)=\Delta_{-}(\beta)+\frac{q_x^2}{2m_x(\beta)}+\frac{q_y^2}{2m_y(\beta)} $
where the masses $ m_x(\beta), m_y(\beta) $ given by
\begin{eqnarray}
    m_x(\beta) & = &  \left \{ \begin{array}{ll}
	2, & \beta\in I  \\
    2\sin2\beta\sqrt{1+\sin^22\beta},  & \beta\in I\!I
    \end{array}     \right.   \nonumber   \\
    m_y(\beta) & = &  \left \{ \begin{array}{ll}
	2/(|\cos2\beta|-\sin^22\beta), & \beta\in I  \\
    \frac{ 2\sin2\beta\sqrt{1+\sin^22\beta} }{ |\cos2\beta|-\sin^22\beta },  & \beta\in  I\!I
    \end{array}     \right.
\label{mass}
\end{eqnarray}
where the regime $ I= (0,\beta_1)\cup(\beta_2,\pi/2) $ and the regime $ I\!I=(\beta_1,\beta_2) $.
The two masses are shown in Fig.4.

\begin{figure}[htb]
\includegraphics[width=6.5cm]{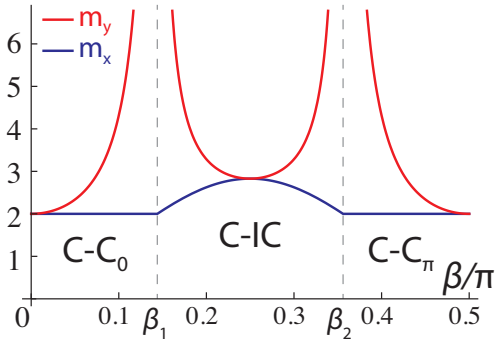}
\caption{(Color online)
The two anisotropic effective masses  $m_y(\beta)\ge m_x(\beta)$ of the magnons.
The equality holds only at $\beta=0,\pi/4,\pi/2$.
$m_y(\beta)$ diverges near the two C-IC boundaries
$m_y( \beta ) \sim |\beta-\beta_{i}|^{-1}$, $i=1,2$. }
\label{finiteT}
\end{figure}

We then obtain the magnetization  $ M(T) $ and specific heat $ C(T) $:
\begin{equation}
\begin{aligned}
	M(T) & = S-\frac{\sqrt{m_x m_y}}{2\pi}Te^{-\Delta/T},    \\
    C(T) & =  \frac{\sqrt{m_x m_y}}{2\pi}(\Delta^2/T) e^{-\Delta/T},
\label{mc}
\end{aligned}
\end{equation}
where $ \Delta=\Delta_{-}(\beta) $ and one can judge the product of the two masses $ m_x m_y $
(or DOS $ D(\epsilon)=\frac{\sqrt{m_x m_y}}{2\pi} \theta( \epsilon-\Delta_{-}) $) is gauge-invariant.
Near $ \beta_1 $ or $ \beta_2 $, the mass $ m_x(\beta) $ is non-critical,
$ m_y( \beta ) \sim | \beta- \beta_1 |^{-1} $ (Fig.4).
It is shown in Sec.V that the $ Y-x $ ground state order at $ (\pi, 0 ) $
and its magnetization $ M(T) $ in Eq.\eqref{mc} are determined
by the sharp peak position and its spectral weight respectively of the equal-time staggered
{\sl longitudinal} spin structure factor $ S^{zz}_{s}( \mathbf{k} ) $ Eq.\eqref{lowzz}.
So both quantities in Eq.\eqref{mc} can be measured by {\sl longitudinal} Bragg spectroscopy
\cite{lightatom1,lightatom2}
and specific heat experiments respectively \cite{heat1,heat2}.

At $ \beta= \beta_1 $ and $ \beta_2 $,
$E_-(\mathbf{q};\beta)=\Delta_{-}(\beta)+\frac{q_x^2}{4}+\frac{q_y^4}{16} $,
Eq.\eqref{mc} should be replaced by
$ M(T)  = S- T^{3/4} e^{-\Delta/T}$,
$C(T) = \Delta^2/T^{\frac{5}{4}} e^{-\Delta/T} $
which implies $ m_y( \beta_1 ) $ can be cutoff at low $T$ as $  m_y( \beta_1 ) \sim T^{-1/2} $.
In fact, at $ \beta= \beta_1 $, the DOS diverges as
$ D(\epsilon)= ( \epsilon-\Delta_{-})^{-1/4} \theta( \epsilon-\Delta_{-}) $.

By adding a uniform magnetic field $-h_u\sum_i  S^{y}_i $ to the Hamiltonian Eq.\eqref{originalH},
following the similar SWE procedures, we can get the expansion of the free energy in terms of $ h_{u} $:
$ F[h_u]= F[0]- \frac{1}{2} \chi_{u} h^{2}_{u} + \cdots $ which leads to the uniform susceptibility:
\begin{eqnarray}
    \chi_{u}(T) =  \left \{ \begin{array}{ll}
	\frac{\sqrt{m_xm_y}}{2\pi} \frac{ m_y | \cos 2 \beta |  }{2} Te^{-\Delta/T},  & \beta\in I
     \\
	\frac{\sqrt{m_xm_y}}{2\pi}( 1- \frac{4}{m^{2}_x} ) e^{-\Delta/T}, & \beta\in I\!I
    \end{array}     \right.
\label{suscu}
\end{eqnarray}

By adding  a $ ( \pi, 0 ) $ staggered magnetic field  $ - h_s \sum_i  (-1)^x S^{y}_i $ to
the Hamiltonian Eq.\eqref{originalH}, the free energy expansion in terms of $ h_{s} $:
$ F[h_s]= F[0]-Mh_s -\frac{1}{2} \chi_{s} h^{2}_{s} + \cdots $ leads to the staggered susceptibility:
\begin{equation}
    \chi_{s}(T)  =  \frac{\sqrt{m_x m_y}}{2\pi} e^{-\Delta/T}
\label{suscs}
\end{equation}

At $\beta= \beta_1 $ and $ \beta_2 $,
one can put $ m_y( \beta_1 ) \sim T^{-1/2} $ in Eq.\eqref{suscu},\eqref{suscs}, one can get
$ \chi_{u}(T) \sim T^{1/4} e^{-\Delta/T}, \chi_{s}(T) \sim T^{-1/4} e^{-\Delta/T} $.

The staggered magnetic field $h_s$ couples to the conserved quantity $ \sum_i  (-1)^x S^{y}_i $,
so can be solely expressed in term of the two effective masses and the gap.
From the specific heat $ C $ in Eq.\eqref{mc} and the staggered susceptibility $ \chi_s $ in Eq.\eqref{suscs},
one can form the Wilson ratio \cite{tqpt,kondo1,kondo2}:
\begin{equation}
	R_w  = \frac{ T \chi_s(T) }{ C(T) }= ( \frac{T}{\Delta } )^2
\label{suscw}
\end{equation}
which only depends on the dimensionless scaling variable of $ T/\Delta $.
Accidentally, it is the same Wilson ratio as that in the $ N_D=4 $ phase in \cite{tqpt}.

New quantum phases and phase transitions
at a finite uniform or a staggered magnetic fields will be mentioned at the conclusion section

\section{ Spin-spin correlation functions at low temperatures }

To directly probe the existence of the C-C$_0$, C-IC, C-C$_{\pi}$ magnons,
one need to evaluate their experimental consequences in spin-spin correlation functions.
For the two sublattice structure $A$ and $B$ (Fig.2a),
one can define\cite{scaling} the uniform spin $ \mathbf{M}=(\mathbf{S}_A+\mathbf{S}_B)/2 $
and the staggered spin $ \mathbf{N}=\mathbf{S}_A-\mathbf{S}_B $.
Then one can define the uniform
$ S^{lm}_u(\mathbf{k},t)= \langle M_l(\mathbf{k},t) M_m(-\mathbf{k},0) \rangle,l,m=1,2,3 $
and staggered
$ S^{lm}_s(\mathbf{k},t)= \langle N_l(\mathbf{k},t) N_m(-\mathbf{k},0) \rangle, l,m=1,2,3 $
spin-spin correlation functions \cite{scaling}.
The spin-orbit coupled U$(1)$ symmetry dictates that there is no mixing between the longitudinal and transverse
components. In the following, one only need to study the uniform and staggered longitudinal and transverse
spin-spin correlation functions separately.

\begin{figure}
\includegraphics[width=6cm]{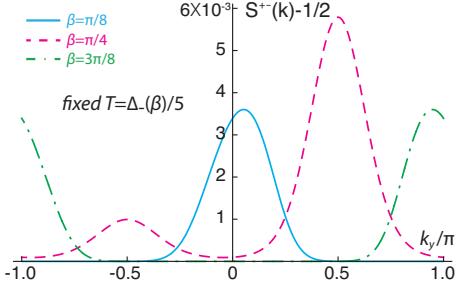}
\caption{(Color online)
The asymmetric shape of the uniform spin structure factor $ S_{u}^{+-}(0, k_y ) $
at the same $ T/\Delta_{-}(\beta) $ 
for C-C$_0$ at $ \beta=\pi/8 $ (blue or solid line), 
C-IC at $ \beta=\pi/4 $ (red or dashed line)
and C-C$_{\pi}$ at $ \beta= 3 \pi/8 $ (green or dash-dot line).
At $ \beta=\pi/8 $, the single peak is slightly shifted from zero to the right due to the spectral weight $ \cos^2 \theta_k/2 $ in Eq.\eqref{equal0}.
At  $ \beta=\pi/4 $, the ratio of two (red or dashed line) Gaussian peak heights located at $ k^{0}_y= \pm \pi/2 $ is $ \frac{ \sqrt{2} + 1 }{ \sqrt{2}-1 } \sim 5.8 $.
At $ \beta=3\pi/8 $, the single peak is slightly shifted from $ \pi $ to the left due to the spectral weight $ \cos^2 \theta_k/2 $ in Eq.\eqref{equal0}.
$ S_{u}^{+-}(0, k_y ) $ can be directly detected by angle resolved transverse atom or light Bragg spectroscopies.
As to be shown in Fig.8 and 9, the asymmetry is eliminated after transforming to the U$(1)$ basis. }
\label{asyorigin}
\end{figure}

\subsection{ Peak positions of the dynamic and Equal-time transverse spin structure factors at low temperatures  }

As shown in appendix C, the spin-orbit coupled U$(1)$ symmetry dictates
the exact relations between the uniform and staggered correlation functions
\begin{equation}
\begin{aligned}
    S_{u}^{+-}(\mathbf{k},\omega) &=  S_{s}^{+-}(\mathbf{k},\omega ), \\
    S_{u}^{++}(\mathbf{k},\omega) &=  -S_{s}^{++}(\mathbf{k},\omega)
\label{usus}
\end{aligned}
\end{equation}
The $ {\cal P}_z $ symmetry dictates that
both $ S_{u}^{+-} $ and $ S_{u}^{+-} $ are even under $ k_x \rightarrow -k_x $.

From Eq.\eqref{energypm}, one can evaluate
the uniform normal and anomalous transverse dynamic spin-spin correlation functions
which has the dimension $ [1/\omega] $:
\begin{equation}
\begin{aligned}
S_{u}^{+-}(\mathbf{k},\omega)  &= 
	\pi \{ \frac{ \sin^2 \frac{\theta_k}{2} }{1-e^{-\omega/T}}
	\left[\delta(\omega\!-\!E_k^+)\!-\!\delta(\omega\!+\!E_k^+) \right]\\
	 & +  	\frac{\cos^2 \frac{\theta_k}{2} }{1-e^{-\omega/T}}
	\left[\delta(\omega\!-\!E_k^-)-\delta(\omega\!+\!E_k^-)\right] \}\\
S_{u}^{++}(\mathbf{k},\omega) & = \frac{\pi}{2}\frac{\sin\theta_k}{1-e^{-\omega/T}}
	\{[\delta(\omega\!-\!E_k^+)-\delta(\omega\!+\!E_k^+)]\\
	 & - 	[\delta(\omega\!-\!E_k^-)-\delta(\omega\!+\!E_k^-)] \}
\label{lowna}
\end{aligned}
\end{equation}
whose poles are given by the excitation spectra $\omega =E_{\pm}(\mathbf{k})$ in Eq.\eqref{energypm}
and the spectral weights are determined by the coefficients of the unitary transformation in Eq.\eqref{unitary}.
Both the excitation spectra and the corresponding spectral weights
in $ S_{u}^{+-}(\mathbf{k},\omega) $ can be measured by
the sharp peak positions of the in-elastic scattering cross sections
of light or atom dynamic {\sl transverse}  Bragg spectroscopy at low temperatures
\cite{lightatom1,lightatom2}.
Unfortunately, $S_{u}^{++}(\mathbf{k},\omega)$ may not be directly measurable.

Due to the gap in the ground state, it is easy to see the normal transverse
susceptibility $ \chi^{+-}(T)= S_{u}^{+-}(\mathbf{k} \rightarrow 0 ,\omega=0)=0 $
and the anomalous transverse susceptibility $ \chi^{++}(T)= S_{u}^{++}(\mathbf{k} \rightarrow 0 ,\omega=0)=0 $.
It is important to observe that the spectral weights in $ S_{u}^{+-}(\mathbf{k},\omega) $ are {\sl not } symmetric
under $ k_y \rightarrow - k_y $,  but  those in $ S_{u}^{++}(\mathbf{k},\omega) $ are.
This is due to the breaking of the $ {\cal P}_x $ and $ {\cal P}_y $ symmetries
of the ground state analyzed in the appendix A.
This is the main difference between the dynamic normal and anomalous spin correlation functions.

\begin{figure}
\includegraphics[width=6cm]{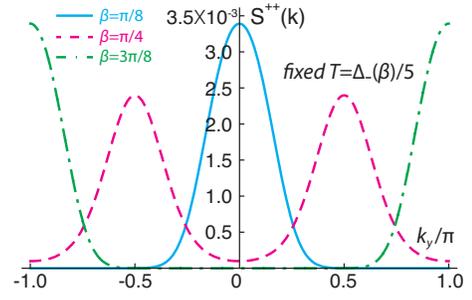}
\caption{(Color online)
The symmetric Gaussian shape of the uniform anomalous spin structure factor $ S_{u}^{++}(0, k_y ) $
at the same $ T/\Delta_{-}(\beta) $ for C-C$_0$ at $ \beta=\pi/8 $, C-IC  at $ \beta=\pi/4 $ and C-C$_{\pi}$ at $ \beta= 3 \pi/8 $.
It becomes a non-Gaussian only near the two C-IC  boundaries.
The Gaussian peak's height and width are determined by the gap in Fig.3
and the effective mass in Fig.4 respectively.
The ratio of the two peak heights [(red or dashed line)/(blue or dash-dot line)] is $ 1/\sqrt{2} $.
Unfortunately, $ S_{u}^{++}(0, k_y ) $ may not be directly detected by
atom or light Bragg spectroscopies. }
\label{splitorigin}
\end{figure}

From above equation, we obtain equal-time spin structure factor
$ S^{lm}_{u,s}( \mathbf{k} )= \int \frac{d \omega}{2 \pi} S^{lm}_{u,s}(\mathbf{k},\omega) $ which is dimensionless:
\begin{equation}
\begin{aligned}
  S_{u}^{+-}(\mathbf{k}) & = \frac{1}{2}
		\frac{\cos^2\frac{\theta_k}{2}}{e^{E_k^-/T}-1}+ \frac{\sin^2\frac{\theta_k}{2}}{e^{E_k^+/T}-1}\\
 S_{u}^{++}(\mathbf{k}) & = 
	\frac{1}{2}\sin\theta_k
	\left(\frac{1}{e^{E_k^-/T}-1}-\frac{1}{e^{E_k^+/T}-1}\right)
\label{equal0}	
\end{aligned}
\end{equation}
where one can see the normal structure factor $ S_{u}^{+-}(\mathbf{k}) $
is {\sl not} symmetric under $ k_y \rightarrow - k_y $,
while the anomalous  $ S_{u}^{++}(\mathbf{k} ) $ is.

One can see that at $ T < T_c \sim \Delta_{-}(\beta ) $ (Fig.3),
the acoustic branch dominates over the optical branch,
then in the regime  $ II=(\beta_1,\beta_2) $, the peak position of
$  S_{u}^{+-}(\mathbf{k}) $ and $ S_{u}^{++}(\mathbf{k}) $
are determined by the two minima positions  $ \mathbf{k}_0 =(0, \pm k^0_y ) $
of the acoustic branch shown in Fig.2b.
As said in Sec.IV, expect at $ \beta_1 ( \beta_2=\pi/2-\beta_1 ) $,
the excitation spectrum can be expanded around the minima $ \mathbf{k}= \mathbf{k}_0+\mathbf{q} $ as
$ E_-(\mathbf{q};\beta)=\Delta_{-}(\beta)+\frac{q_x^2}{2m_x(\beta)}+\frac{q_y^2}{2m_y(\beta)} $
where the masses $m_x(\beta)$, $m_y(\beta)$
are given above Eq.\eqref{mc}. 
We reach simplified and physically transparent expressions:
\begin{equation}
\begin{aligned}
  S_{u}^{+-}(\mathbf{k}) & \sim   \frac{1}{2}
	\!+\!	\cos^2\!\frac{\theta_k}{2}  e^{-\frac{\Delta_{-}(\beta)}{T}} e^{- ( \frac{q_x^2}{2m_x(\beta)}+\frac{q_y^2}{2m_y(\beta)})/T }   \\
 S_{u}^{++}(\mathbf{k}) & \sim  
	\frac{1}{2}\sin\theta_k e^{-\frac{\Delta_{-}(\beta)}{T}} e^{- ( \frac{q_x^2}{2m_x(\beta)}+\frac{q_y^2}{2m_y(\beta)})/T }
\label{equal1}	
\end{aligned}
\end{equation}
where $ \mathbf{k} $ belongs to reduced Brillouin zone (RBZ).
At the two C-IC boundaries $ \beta=\beta_1, \beta_2 $,
it becomes a non-Gaussian $ \sim e^{-\frac{ q^4_{y}}{16 m_y(\beta) T}} $.

Because the peak splitting process only happens
in the $ k_y $ axis, so we only show
$ S_{u}^{+-}(\mathbf{k} ) $ and $ S_{u}^{++}(\mathbf{k} ) $ at $ k_x=0 $ in Fig.5 and Fig.6 respectively.
Along the $ k_x $ axis, it is a  Gaussian peak with the width $ \sigma_x= \sqrt{ m_{x}(\beta) T } $.
In fact, when drawing the Fig.5 and Fig.6, we used the Eq.\eqref{equal0} where we took the complete
expression Eq.\eqref{energypm} for $ E_k^- $ and dropped the optical branch.
We also drew the same figure using Eq.\eqref{equal1} and found very little difference at several temperatures
$ T/\Delta_{-}(\beta) =1/2,1/3,1/5,1/10 $, so Eq.\eqref{equal1} is quite accurate.

Shown in Fig.5 is $ S_{u}^{+-}(\mathbf{k} ) $.
At C-C$_0$ regime, the asymmetric peak is pinned slightly right to  $ (0,0) $.
At C-C$_{\pi}$ regime, the asymmetric peak is pinned slightly left to $ (0,\pm \pi) $.
At C-IC regime, the peak splits into two Gaussian peaks
located at $ (0,\pm k^{0}_y ) $ continuously tuned by the SOC strength.
Well inside the C-IC regime, the two Gaussian peaks
have the heights $ \frac{1}{2} (1+ \cos\theta_{\pm k^{0}_y} ) e^{-\Delta_{-}(\beta)/T} $
and the same width along the $ k_y $ axis $ \sigma_y= \sqrt{ m_{y}(\beta) T } $.
Due to asymmetry under $ k_y \rightarrow - k_y $,
the ratio of the two Gaussian peaks is $ (1+ \cos\theta_{k^{0}_y} )/(1+ \cos\theta_{-k^{0}_y} ) $.
At $ \beta= \pi/4, k^{0}_y=\pi/2 $, 
the ratio becomes $ \frac{ \sqrt{2} + 1 }{ \sqrt{2}-1 } \sim 5.8 $.
So the ratio of the two peak heights, the heights and their widths are
effective measures of the unitary transformation Eq.\eqref{unitary}, the gap and the effective mass respectively.
All these features can be directly measured by the angle
resolved light or atom {\sl transverse}  Bragg spectroscopy at low temperatures
\cite{lightatom1,lightatom2}.
The C-IC has a larger gap at the center in Fig.3,
so can be more easily detected than C-C$_0$ and C-C$_{\pi}$.
The two split Gaussian peaks driven by C-IC magnons
in the {\sl transverse} spin structure factors $ S_{u}^{+-}(\mathbf{k} ) $
is a unique and salient feature of the RH model.

Shown in Fig.6 is $ S_{u}^{++}(\mathbf{k}) $.
At C-C$_0$ regime, the Gaussian peak is pinned at  $ (0,0) $.
The height and width of the Gaussian peak is given in Eq.\eqref{equal1}.
At C-C$_{\pi}$ regime, the  peak is pinned  at $ (0,\pm \pi) $.
At the C-IC regime, the peak splits into two Gaussian peaks
located at $ (0, \pm k^{0}_y ) $ continuously tuned by the SOC strength.
They have the same height $ \frac{1}{2} \sin\theta_k e^{-\Delta_{-}(\beta)/T} $
and the width   $ \sigma_y= \sqrt{ m_{y}(\beta) T } $.
The ratio of the peak height at the I-IC point over that at the C-C$_0$ (or C-C$_{\pi}$) point
is given by $ \sin\theta_{k_0}  < 1 $.
At $ \beta= \pi/4, k^{0}_y= \pi/2 $, the ratio becomes $ 1/\sqrt{2} $.
So the ratio of the peak heights,
the height itself and its width are effective measures of the unitary transformation,
the gap and the effective mass respectively.
Unfortunately, $ S_{u}^{++}(\mathbf{k}) $ may not be directly measurable by the Bragg spectroscopy.

\subsection{ Longitudinal spin correlation functions: Ground state and magnetization detection}

One can also evaluate the uniform and staggered {\sl connected } dynamic longitudinal spin-spin correlation functions at low temperatures:
\begin{widetext}
\begin{equation}
\begin{aligned}
	S_{u}^{zz}(\mathbf{k},\omega)  =
	\frac{2\pi}{N}\sum_q
	\left\{
		\cos^2\frac{\theta_q+\theta_{q+k}}{2}
		[n_q^+(1+n_{q+k}^+)\delta(\omega+E_q^+-E_{q+k}^+)
		+n_q^-(1+n_{q+k}^-)\delta(\omega+E_q^--E_{q+k}^-)]
	\right.\\
	\left.
		+\sin^2\frac{\theta_q+\theta_{q+k}}{2}
		[n_q^+(1+n_{q+k}^-)\delta(\omega+E_q^+-E_{q+k}^-)
		+n_q^-(1+n_{q+k}^+)\delta(\omega+E_q^--E_{q+k}^+)]
	\right\}   \\
	S_{s}^{zz}(\mathbf{k},\omega)  =
	\frac{2\pi}{N}\sum_q
	\left\{
		\cos^2\frac{\theta_q-\theta_{q+k}}{2}
		[n_q^+(1+n_{q+k}^+)\delta(\omega+E_q^+-E_{q+k}^+)
		+n_q^-(1+n_{q+k}^-)\delta(\omega+E_q^--E_{q+k}^-)]
	\right.  \\
	\left.
		+\sin^2\frac{\theta_q-\theta_{q+k}}{2}
		[n_q^+(1+n_{q+k}^-)\delta(\omega+E_q^+-E_{q+k}^-)
		+n_q^-(1+n_{q+k}^+)\delta(\omega+E_q^--E_{q+k}^+)]
	\right\}
\label{lowzz}
\end{aligned}
\end{equation}
\end{widetext}
which include both the intra-band transitions and the inter-band transition
between the optical $E_k^+$  and the  acoustic  $E_k^-$.

It is easy to see that due to the summation over the momentum transfer in Eq.\eqref{lowzz},
so the dynamic connected longitudinal spin-spin correlation functions will just show a broad distribution,
in sharp contrast to the transverse dynamic correlation functions Eq.\eqref{lowna}.
One can also evaluate the uniform  $ \chi_u(T)= S_{u}^{zz}( \mathbf{k} \rightarrow 0, \omega=0) $
and $ ( \pi, 0 ) $ staggered susceptibility $ \chi_s(T)= S_{s}^{zz}( \mathbf{k} \rightarrow 0, \omega=0) $
and reproduce the results in Eq.\eqref{suscu} and \eqref{suscs} respectively.

The equal-time longitudinal spin structure factors follow
$ S_{u,s}^{zz}(\mathbf{k}) = \int \frac{d \omega}{2 \pi} S_{u,s}^{zz}(\mathbf{k},\omega) $:
\begin{widetext}
\begin{equation}
\begin{aligned}
   S_{u}^{zz}(\mathbf{k})&=
	\frac{1}{N}\sum_q
	\left\{
		\cos^2\frac{\theta_q\!+\!\theta_{q+k}}{2}
		[n_q^+(1\!+\!n_{q+k}^+)+n_q^-(1\!+\!n_{q+k}^-)]
		+\sin^2\frac{\theta_q\!+\!\theta_{q+k}}{2}
		[n_q^+(1\!+\!n_{q+k}^-)+n_q^-(1\!+\!n_{q+k}^+)]
	\right\}   \\
	S_{s}^{zz}(\mathbf{k})&=
	\frac{1}{N}\sum_q
	\left\{
		\cos^2\frac{\theta_q\!-\!\theta_{q+k}}{2}
		[n_q^+(1\!+\!n_{q+k}^+)+n_q^-(1\!+\!n_{q+k}^-)]
		+\sin^2\frac{\theta_q\!-\!\theta_{q+k}}{2}
		[n_q^+(1\!+\!n_{q+k}^-)+n_q^-(1\!+\!n_{q+k}^+)]
	\right\}
\label{lowequal}
\end{aligned}
\end{equation}
\end{widetext}
which, at low temperatures $ T < \Delta_{-}( \beta ) $, can be simplified to:
\begin{equation}
\begin{aligned}
   S_{u}^{zz}(\mathbf{k}) & = 
	\frac{1}{N}\sum_q  n_q 
	+ \frac{1}{N}\sum_q \cos^2\frac{\theta_q\!+\!\theta_{q+k}}{2} n_q^- n_{q+k}^- + \cdots\\
	S_{s}^{zz}(\mathbf{k}) & =   
	\frac{1}{N}\sum_q n_q 
	+ \frac{1}{N}\sum_q \cos^2\frac{\theta_q\!-\!\theta_{q+k}}{2} n_q^- n_{q+k}^- + \cdots
\label{lowzzequalleading}	
\end{aligned}
\end{equation}
where $ n_q=n_q^{+} + n_q^{-} $ and $ \cdots $ mean the sub-leading terms at low temperatures.
Again, due to the summation over the momentum transfer in Eq.\eqref{lowzzequalleading},
so the longitudinal spin structure factors will just show a broad distribution,
in sharp contrast to the transverse spin structure factors in Eq.\eqref{equal0}.

Note that in the staggered {\sl connected } dynamic (equal-time) longitudinal spin-spin correlation function
$ S_{s}^{zz}(\mathbf{k},\omega) $ ($ S_{s}^{zz}(\mathbf{k}) $)
in Eq.\eqref{lowzz} ( in Eq.\eqref{lowequal} ),
we have subtracted the magnetization part
$  M^{2}(T) \delta_{ \mathbf{k},0} 2 \pi \delta( \omega ) $
($ M^{2}(T) \delta_{ \mathbf{k},0} $)
due to the symmetry breaking \cite{direct} in the quantum  ground state in Fig.2a.
The magnetization $ M(T) $ is given by Eq.\eqref{mc}.
The symmetry breaking and the magnetization can be detected by the sharp peak at momentum
$ (\pi,0 ) $ ($ (0,0) $ in the RBZ)
and its spectral weight of the longitudinal Bragg spectroscopy at low temperatures
\cite{lightatom1,lightatom2}.

\section{ Specific heat and spin structure factors at high temperatures }

It was known that the spin wave expansion only works at low temperature $ T \ll T_c $.
At $ T > T_c $, the magnetization vanishes, all the symmetries of the Hamiltonian Eq.\eqref{originalH} analyzed in appendix A were restored,
so there is no $ A $ and $ B $ structure anymore.
At high temperatures  $ T \gg T_c $, one need to use the high temperature expansion by expanding the spectral weight
$ e^{-H/T}=\sum_{n=0}^{\infty}\frac{(-1)^n}{n!}\frac{H^n}{T^n} $. In this section, we focus on $ S=1/2 $.

\subsection{ Specific heat and Wilson loop detections }

    We also obtain the high temperature expansion of the specific heat per site to the order of $  (J/T)^{4} $:
\begin{equation}
	C(T)/N=\frac{3}{8} (\frac{J}{T})^2-\frac{3}{16} ( \frac{J}{T} )^3
	+\frac{12\cos4\beta-33}{128} ( \frac{J}{T} )^4
\label{hightheat}
\end{equation}
    which depends on $ \beta $ starting at the order of $  (J/T)^{4} $. Obviously, at the two Abelian points $ \beta=0, \pi/2 $,
    it recovers that of the Heisenberg model to the same order, reaches the minimum at the most frustrated point $ \beta=\pi/4 $ (Fig.1b).
    It is important to observe that $ \cos 4 \beta $ is nothing but the Wilson loop around a unit cell $  \cos 4 \beta=\frac{ W_R-1 }{2} $.
    We expect that the whole high temperature expansion series of the specific heat $ C_v/N $ can be expressed in terms of the whole set
    of Wilson loops order by order in $ \frac{J}{T} $.
    This set-up the principle that the whole set of Wilson loops with $ n $ edges in the RH can be experimentally measured
    at the corresponding orders of $ (J/T)^{n} $ by specific heat measurements \cite{heat1,heat2}.

\subsection{  Equal-time transverse spin structure factors at high temperatures }

   At $ T > T_c $, because all the symmetries of the Hamiltonian Eq.\eqref{originalH} were restored, so there is no $ A $
and $ B $ structure anymore. We get the equal-time normal and anomalous transverse spin-structure factors
to the order of $  (J/T)^{2} $:
\begin{equation}
\begin{aligned}
S^{+-}(\mathbf{k})  
	=\left(\frac{J}{4T}-\frac{J^2}{16T^2}\right)\cos(k_y+2\beta)\\
	  +\frac{J^2}{16T^2}[\cos 2k_x+\cos(2k_y+4\beta)]   \\
S^{++}(\mathbf{k})
	=\left(\frac{J}{4T}-\frac{J^2}{16T^2}\right)\cos k_x \\
      	  +\frac{J^2}{8T^2} \cos2\beta  [\cos (k_x+k_y)+\cos(k_x-k_y)]
\label{hightna}
\end{aligned}
\end{equation}
    where the  explicit dependence on the gauge parameter $ \beta $ in $ S^{+-}(\mathbf{k}) $
    can be easily detected by  angle-resolved {\sl transverse } light or atom Bragg scattering experiments \cite{lightatom1,lightatom2}.
    Again, one can observe that $ S_{u}^{+-}(\mathbf{k} ) $ is {\sl not} symmetric under $ k_y \rightarrow - k_y $,
    but $ S_{u}^{++}(\mathbf{k},\omega) $ is.

   In order to make comparisons with the low temperature expressions Eq.\eqref{equal0}, also contrast with the corresponding
   expressions in the U$(1)$ basis to be discussed in Sec.VIII, we split Eq.\eqref{hightna} into sublattice A and B
   in Fig.2a,  then form a uniform and staggered spin structure factors:
\begin{equation}
\begin{aligned}
S_{u}^{+-}(\mathbf{k}) 
	=\left(\frac{J}{4T}-\frac{J^2}{16T^2}\right)\cos(k_y+2\beta)\\
	  +\frac{J^2}{16T^2}[\cos 2k_x+\cos(2k_y+4\beta)],\\
S_{u}^{++}(\mathbf{k})  
	=\left(\frac{J}{4T}-\frac{J^2}{16T^2}\right)\cos k_x\\
	  +\frac{J^2}{8T^2} \cos2\beta  [\cos (k_x+k_y)+\cos(k_x-k_y)]	
\label{hightnasplit}
\end{aligned}
\end{equation}
which will be compared to those in the U$(1)$ basis in the Sec.VIII.

\subsection{ Longitudinal spin structure factor at high temperatures}

  One can also evaluate the equal-time longitudinal spin structure factor at high temperatures:
\begin{align}
	S^{zz}(\mathbf{k})
	=&\left(-\frac{J}{8T}+\frac{J^2}{32T^2}\right)[\cos k_x-\cos k_y]\nonumber\\
	&+\frac{J^2}{32T^2}[\cos2k_x+\cos2k_y]\nonumber\\
	&-\frac{J^2}{16T^2}[\cos(k_x+k_y)+\cos(k_x-k_y)]
\label{hightl}
\end{align}
   which is independent of $ \beta $ to the order of $ (J/T)^2 $.
   In fact, it can be shown that Eq.\eqref{hightl} coincides with that of the Heisenberg model to the same  order.
   However, we expect the $ \beta $ dependence will appear  in the order of  $ (J/T)^4 $.

   In order to make comparisons with the low temperature expressions Eq.\eqref{lowzz}, also contrast with the corresponding
   expressions in the U$(1)$ basis to be discussed in the section VIII, we split Eq.\eqref{hightl} into sublattice A and B
   in Fig.2a,  then form a uniform and staggered spin structure factors:
\begin{widetext}
\begin{equation}
\begin{aligned}
S_{u}^{zz}(\mathbf{k}) & = 
	\left( -\frac{J}{8T} + \frac{J^2}{32T^2}\right)
	(\cos k_x - \cos k_y)
	+\frac{J^2}{32T^2}
	[\cos 2k_x+\cos 2k_y]
	-\frac{J^2}{16T^2}
	[\cos (k_x+k_y)+\cos (k_x-k_y)]       \\
S_{s}^{zz}(\mathbf{k}) & = 
	\left(\frac{J}{8T}-\frac{J^2}{32T^2}\right)
	(\cos k_x+\cos k_y)
	+\frac{J^2}{32T^2}
	[\cos 2k_x+\cos 2k_y]
	+\frac{J^2}{16T^2}
	[\cos (k_x+k_y)+\cos (k_x-k_y)]
\label{hightlsplit}
\end{aligned}
\end{equation}
\end{widetext}
which will be compared to those in the U$(1)$ basis below.

\section{ Experimental realizations of the RH models in the U$(1)$ basis }

 By a local gauge transformation $ \tilde{b}_i=(i\sigma_x)^{i_x}b_i $ on Eq.\eqref{bosonint} along the dashed line in Fig.1b
 to get rid of the gauge fields on all
 the $x-$links, then a global rotation  $ \tilde{\tilde{b}}_i=e^{-i\frac{\pi}{4}\sigma_x}\tilde{b}_i $ to rotate $ S^{y} $ to $ S^{z} $,
 Eq.\eqref{bosonint} becomes:
\begin{eqnarray}
 \tilde{\tilde{H}}_{U(1)} & = & -t\sum_{i} [
	\tilde{\tilde{b}}_i^\dagger \tilde{\tilde{b}}_{i+x}
	+\tilde{\tilde{b}}_i^\dagger e^{(-1)^{i_x} i\beta\sigma_z} \tilde{\tilde{b}}_{i+y}+h.c.]
   \nonumber   \\
   &+ & \frac{U}{2} \sum_{i} ( \tilde{\tilde{n}}_{i}-N)^2
\label{kinetic1}
\end{eqnarray}
  where all the remaining gauge fields on the $y-$links commute.
  Obviously, the spin-orbital coupled U$(1)$ symmetry in the original basis Eq.\eqref{bosonint} becomes explicit
  in this ``U$(1)$'' basis with the conserved quantity
  $\sum \tilde{\tilde{S}}_i^z=\sum \tilde{S}_i^y=\sum(-1)^{i_x}S_i^y$.
  In Fig.7, we contrast the gauge field configurations in the U$(1)$ basis with that quantum spin Hall effect realized in recent
 experiments \cite{uniform1,uniform2,uniform3}.

\begin{figure}[htb]
\includegraphics[width=8.5cm]{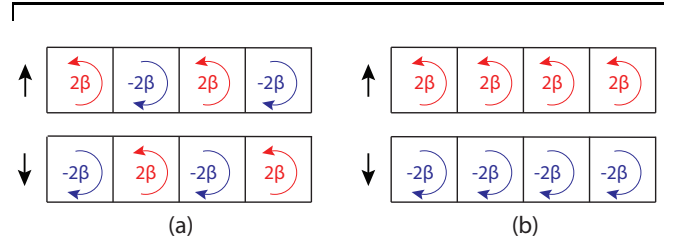}
\caption{(Color online)
Gauge fields  in (a) in the U$(1)$ basis in Eq.\eqref{kinetic1}.
(b) Quantum spin Hall Hamiltonian realized in recent experiments \cite{uniform1,uniform2,uniform3,honey1,dual1,dual2,dual3}.
 Both are translational invariant along the $ y $ direction, so only one row is shown. }
\label{fig5}
\end{figure}

  A specific experimental implementation scheme for the U$(1)$ basis  in Fig.7a can be suggested in the following.
  We first introduce the anisotropy $ \lambda $ in the interaction term in Eq.\eqref{kinetic1}:
\begin{equation}
  V_{int}( \lambda )
	=\frac{U}{2} \sum_{i} ( n^2_{i \uparrow}
		+n^2_{i \downarrow}
		+2\lambda n_{i \uparrow} n_{i \downarrow} )
\label{anisotropy}
\end{equation}
 To keep at the integer filling $ N $, the chemical potential will also be adjusted accordingly.
   Now if setting $ \lambda=0 $ and the chemical potential $ \mu(\lambda=0)= U N/2 $ to keep the total filling at $ \langle n \rangle = N $,
   the interaction term becomes:
\begin{equation}
    V_{int}(\lambda=0)
	=\frac{U}{2} \sum_{i} [ ( n_{i \uparrow}-N/2)^2 +  ( n_{i \downarrow}-N/2)^2 ]
\label{decouple}
\end{equation}
   where each spin spin species occupies half integer fillings $ N/2 $.
   Then Eq.\eqref{kinetic1} decouples into two identical copies of spin up and spin down, each is in the SF state for $ N=1 $ for all $ U $
   ( we set $ N=1 $ in the following ).
   For the spin up, the magnetic field is $ \pm 2 \beta $  alternating along $ x $ direction, for spin down, the magnetic field
 is just reversed to keep the Time reversal symmetry. So for the spin up, the staggered magnetic field can be
 realized in the previous experiments
 \cite{stagg1,stagg2,stagg3,stagg4}.
 For  the spin down, as demonstrated in \cite{uniform3}, if it carries opposite magnetic moment to the spin up state, then it will
 experience the opposite magnetic field.
 Now one can adiabatically turn on the inter-species interaction between the two pseduo-spin components
 $ 2 \lambda  n_{i \uparrow} n_{i \downarrow} $, setting $ \lambda =1 $ will recover Eq.\eqref{kinetic1}.
 The two pseduo-spin components cab be  two suitably chosen hyperfine states for $^{87}$Rb  or the two isotopes of the highly magnetic element
 dysprosium \cite{dy}: $ ^{162}$Dy  and $ ^{160}$Dy .
 Note that by turning on $ \lambda $ this way, the U$(1)$ symmetry is kept for all $ \lambda $ (see Fig.10).
 The dramatic effects of the spin anisotropic interaction $ 0 < \lambda < 1 $ will be presented elsewhere \cite{un1}.
 Obviously, in the strong coupling limit $ U \gg t $, as $ \lambda $ increases, the system will evolve from the SF state at small $ \lambda $
 to the $Y$-$x$ state at $ \lambda=1 $.
 We conclude that the U$(1)$ basis could be realized in some combination of previous experiments to realize staggered magnetic field \cite{stagg1,stagg2} and recent experiments to realize quantum spin Hall effects \cite{uniform1,uniform2,uniform3}.

  The quantum spin Hall Hamiltonian corresponding to Fig.7b is:
\begin{eqnarray}
    H_{QSH} & = & -t\sum_{i} [b_i^\dagger b_{i+x}
	+ b_i^\dagger e^{ i 2 \beta x \sigma_z} b_{i+y}+h.c.]
         \nonumber  \\
    & + & \frac{U}{2} \sum_{i} ( n_{i} - N )^2
\end{eqnarray}
 where the $ x $ is the $ x-$ coordinate
\cite{honey1,dual1,dual2,dual3}
 of the site $ i $. For irrational $ \beta $, this Hamiltonian completely breaks the lattice translational symmetry.
 For a rational $ 2 \beta=p/q $, it contains $ q $ sites per unit cell (RBZ is $ 1/q $ of the original BZ, for details, see \cite{honey1,dual1,dual2,dual3}).
 However, the U$(1)$ basis Fig.7a only breaks the lattice into $ A $ and $ B$ sublattices  for any ( irrational ) value $ \beta $.
 So the two Hamiltonian are dramatically different.

 As pointed out in \cite{uniform3},  non-Abelian gauge in Eq.\eqref{bosonint}
 can be achieved by adding spin-flip Raman lasers to induce a $ \alpha \sigma_x $ term along the
 horizontal bond in Fig.1a, or by driving the spin-flip transition with RF or microwave fields.
 If so, the original basis can also be realized in near future experiments.

  It was known \cite{ho} that for $  V_0/E_r \ge 10 $ where $ V_0 $ is the optical lattice potential and $ E_r $ is the recoil energy,
  the spinor boson Hubbard model Eq.\eqref{bosonint} is well within the strong coupling regime $ J \ll t \ll U $ .
  For $^{87}$Rb  atoms used in the recent experiments \cite{uniform1,uniform2,uniform3}, the superfluid-insulator transition is estimated to be
  $ V_0/E_r \sim 12 $, so the RH model Eq.\eqref{rh} applies well in the regime.
  Near the most frustrated point $ \beta = \pi/4 $, the critical temperature $ T_c \sim J \sim
  0.2 nK $. It remains experimentally challenging to reach such low temperatures\cite{ho}. However, in view of two recent
  advances of new cooling techniques \cite{cool1,cool2} to reach $ 0.35 nK $,
  the obstacles maybe overcame in the near future.
  Before reaching such low temperatures, the specific heat measurement \cite{heat1,heat2} at high temperatures to
  determine the whole sets of Wilson loops order by order in $ J/T $
  along the dashed line in Fig.1b could be performed easily.

  Because the U$(1)$ basis can be realized in current experiments, so it is important to work out
  various experimental measurable quantities in this basis explicitly.
  As first stressed in \cite{tqpt} that
  in contrast to condensed matter experiments where only gauge invariant
  quantities can be measured, both gauge invariant and non-gauge invariant quantities can be measured by experimentally generating various non-Abelian gauges corresponding to the same set of Wilson loops.
  Some quantities such as the absolute value of the magnetization $ M(T) $, specific heat $ C_v $, the
  gaps and density of states are gauge invariant, so are the same in both basis. The uniform $ \chi_u $ and the staggered
  susceptibilities $ \chi_s $ will exchange their roles between the original and the U$(1)$ basis.
  However, the spin-spin correlations functions are gauge
  dependent \cite{tqpt}, so will be explicitly computed at both low and high temperatures in the next section.
  We will also comment on the nature of the finite temperature phase transition in Fig.3.

\section{ Spin-spin correlation functions in the U$(1)$ basis }

We first make a local rotation $ \tilde{\mathbf{S}}_n = R( \hat{x}, \pi n_1 ) \mathbf{S}_n $ to get rid of the $R$-matrix on the $x$-links in Fig.1a,
then just as in the original basis, we make a global rotation \cite{direct} $ \tilde{\tilde{\mathbf{S}}}_n = R_x( \pi/2 ) \tilde{\mathbf{S}}_n $
to rotate the spin quantization axis from $ Y $ to $ Z $, we reach the Hamiltonian in the U$(1)$ basis \cite{drop}:
\begin{widetext}
\begin{eqnarray}
	H_{U(1)} &=&-J\sum_{i\in A}
	[\frac{1}{2}(S_i^+S_{i+x}^-+S_i^-S_{i+x}^+)+S_i^zS_{i+x}^z
	+\frac{1}{2}(e^{i2\beta}S_i^+S_{i+y}^-+e^{-i2\beta}S_i^-S_{i+y}^+)
	+S_i^zS_{i+y}^z] \nonumber\\
	&-&J\sum_{j\in B}
	[\frac{1}{2}(S_j^+S_{j+x}^-+S_j^-S_{j+x}^+)+S_j^zS_{j+x}^z
	+\frac{1}{2}(e^{-i2\beta}S_j^+S_{j+y}^-+e^{i2\beta}S_j^-S_{j+y}^+)
	+S_j^zS_{j+y}^z]
\label{u1h}
\end{eqnarray}
\end{widetext}
    where $ A $ and $ B $ are the two sublattices in Fig.2a.

    By comparing with the Hamiltonian in the original basis Eq.\eqref{originalH}, we can see that in the U$(1)$ basis,
    due to the absence of the anomalous terms like $ S^{+}S^{+} $ or $  S^{-}S^{-} $,
    the U$(1)$ symmetry with the conservation $ \sum_i S^{z}_{i} $ is explicit, but at the expense of
    the translational symmetry explicitly broken due to the local spin rotation $ \tilde{\mathbf{S}}_n = R( \hat{x}, \pi n_1 ) \mathbf{S}_n $.
    It is easy to see the exact ground state $Y$-$x$ in Fig.2a in the original basis becomes simply a Ferromagnetic state
    along $ Y $ direction in the U$(1)$ basis.

    The symmetry of Eq.\eqref{u1h} can be obtained by performing the local gauge transformation on the symmetries in the original
    basis analyzed in the appendix A.

\subsection{ Spin-spin correlation functions at low temperatures: Spin wave expansions}

     In the U$(1)$ basis Eq.\eqref{u1h}, introducing two sets of HP bosons $  S^{+}=  \sqrt{2S-a^{\dagger} a }a, S^{-}= a^{\dagger} \sqrt{2S-a^{\dagger} a },
     S^{z}=S- a^{\dagger} a $ for the sublattice A and $  S^{+}=  \sqrt{2S-b^{\dagger} b }b, S^{-}= b^{\dagger} \sqrt{2S-b^{\dagger} b },
     S^{z}=S- b^{\dagger} b $ for the sublattice B \cite{note}, we find the Hamiltonian in terms of the HP bosons becomes identical to
     that in the original basis, so the excitation spectra in Eq.\eqref{energypm} follow , the unique and salient features of the
     C-C$_0$, C-IC, C-C$_{\pi}$, the gaps of the acoustic and optical branches shown in Fig.2b, Fig.3 and Fig.4 remain the same in the U$(1)$ basis.
     However, as to be shown below, Fig.5 will be replaced by Fig.8 and 9.

{\sl 1. Sharp peak positions of dynamic transverse spin-spin correlation functions: excitation spectra }

     Note that the U$(1)$ symmetry dictates that there is no anomalous spin-spin correlation functions:
\begin{equation}
        S_{u,U(1)}^{++}(\mathbf{k},\omega)=S_{s,U(1)}^{++}(\mathbf{k},\omega)=0
\label{u1zero}
\end{equation}

     Using the HP bosons, we find the uniform and staggered transverse dynamic spin-spin correlation functions:
\begin{equation}
\begin{aligned}
	S_{u,U(1)}^{+-}(\mathbf{k},\omega) & = 
	\pi
	[	\frac{1-\sin\theta_k}{1-e^{-E_k^+/T}}\delta(\omega-E_k^+)
         \\
		& +   \frac{1+\sin\theta_k}{1-e^{-E_k^-/T}}\delta(\omega-E_k^-) ] \\
	S_{s,U(1)}^{+-}(\mathbf{k},\omega) & = 
	\pi
	[	\frac{1+\sin\theta_k}{1-e^{-E_k^+/T}}\delta(\omega-E_k^+)\\
		& +   \frac{1-\sin\theta_k}{1-e^{-E_k^-/T}}\delta(\omega-E_k^-) ]
\label{lowpmu1}
\end{aligned}
\end{equation}
      which is indeed different from Eq.\eqref{lowna} in the original basis.
      Both $ S_{u,U(1)}^{+-}(\mathbf{k},\omega) $ and
      $ S_{s,U(1)}^{+-}(\mathbf{k},\omega) $ are symmetric under the space inversion $ \mathbf{k} \rightarrow - \mathbf{k} $.
      As shown in the appendix D, performing the local spin
      rotation  $ \tilde{\mathbf{S}}_n = R( \hat{x}, \pi n_1 ) \mathbf{S}_n $ on Eq.\eqref{lowpmu1} does not lead to Eq.\eqref{lowna}.

   Both the uniform and staggered transverse dynamic spin-spin correlation functions in Eq.\eqref{lowpmu1} can be
   easily detected by light or atom Bragg scattering experiments \cite{lightatom1,lightatom2}. So
   both the optical $ E_k^+ $  and the  acoustic  $ E_k^- $ excitation spectra can be extracted from the peak positions of scattering cross
   sections of these experiments.  Taking $ T \rightarrow 0 $ limit in Eq.\eqref{lowpmu1}, we can see that the DOS of the spin excitations is given by
\begin{equation}
      D(\omega)= \int \frac{ d^{2} \mathbf{k} }{ (2 \pi)^{2} }  [ S_{u,U(1)}^{+-}(\mathbf{k},\omega) + S_{s,U(1)}^{+-}(\mathbf{k},\omega) ]
\label{dos}
\end{equation}
      which can be detected by energy Bragg spectroscopy \cite{lightatom1,lightatom2}
      or more directly by the {\sl In-Situ } measurements \cite{dosexp}.

\begin{figure}
\includegraphics[width=6cm]{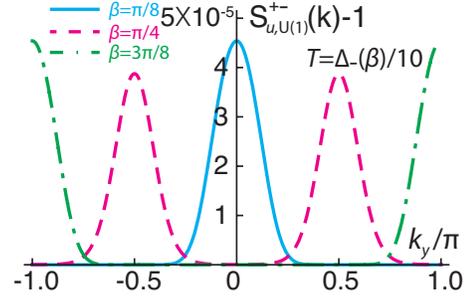}
\caption{(Color online)
The uniform spin structure factor $ S_{u,U(1)}^{+-}(\mathbf{k}) $ at  $ T/\Delta_{-}(\beta) =1/10 $
for C-C$_0$ at $ \beta=\pi/8 $,  
C-IC  at $ \beta=\pi/4 $ and 
C-C$_{\pi} $ at $ \beta= 3 \pi/8 $.
The ratio of the two peak heights [(red or dashed line)/(blue or dash-dot line)] is $ \frac{1}{2}(1 + \frac{1}{\sqrt{2}}) $.
By comparing with Fig.5 in the original basis, one can see the asymmetry is eliminated. 
So C-C$_0$, C-C$_{\pi}$ and C-IC can be
more easily distinguished in the U$(1)$ basis than in the original basis.
As shown in \cite{zeeman}, another advantage is that it is much more easier to determine the spin-orbital structures of
possible phases when the system is subject to a Zeeman field or a spin-anisotropic interaction respecting the U$(1)$ symmetry.  }
\label{splitu10}
\end{figure}

\begin{figure}
\includegraphics[width=6cm]{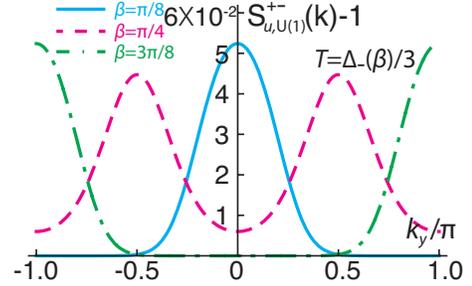}
\caption{ The uniform spin structure factor $ S_{u,U(1)}^{+-}(\mathbf{k}) $ at  $ T/\Delta_{-}(\beta) =1/3 $.
It is instructive to compare with Fig.5 in the original basis. }
\label{splitu11}
\end{figure}

{\sl 2. Gaussian peak positions of equal-time transverse spin structure factors: 
C-C$_0$,C-C$_{\pi}$ and C-IC magnons. }

The equal time spin structure factors 
$ S^{+-}_{u,s;U(1)}( \mathbf{k} )
= \int \frac{d \omega}{2 \pi} S^{+-}_{u,s;U(1)}(\mathbf{k},\omega) $  follow:
\begin{equation}
\begin{aligned}
  S_{u,U(1)}^{+-}(\mathbf{k}) & = &  1+
		\frac{1}{2} (\frac{1 + \sin \theta_k }{e^{E_k^-/T}-1}+ \frac{ 1-\sin \theta_k }{e^{E_k^+/T}-1})\\
  S_{s,U(1)}^{+-}(\mathbf{k}) & = &  1+\frac{1}{2} (
		\frac{1 - \sin \theta_k }{e^{E_k^-/T}-1}+ \frac{ 1+\sin \theta_k }{e^{E_k^+/T}-1})
\label{equal0u1}
\end{aligned}	
\end{equation}
     where one can see both $ S_{u,U(1)}^{+-}(\mathbf{k} ) $ and
     $ S_{s,U(1)}^{+-}(\mathbf{k} ) $ are symmetric under the space inversion $ \mathbf{k} \rightarrow - \mathbf{k} $.
     However, the uniform structure factor $ S_{u,U(1)}^{+-}(\mathbf{k}) $ has a higher spectral weight
     $ 1 + \sin \theta_k $ on the acoustic branch, lower one $   1 - \sin \theta_k $ on the optical branch,
     the staggered structure factor $ S_{s,U(1)}^{+-}(\mathbf{k}) $ is just opposite. So  the uniform structure factor
     is a better quantity to measure the acoustic branch by the Bragg spectroscopy. Of course, it is also a easier one
     to measure than the staggered structure factor.

     Similar manipulations following  Eq.\eqref{equal0}
     apply here also. Shown in Fig.8 and 9 are the uniform spin structure factor $ S_{u,U(1)}^{+-}(\mathbf{k}) $ at two different temperatures.
     We conclude that in the U$(1)$ basis, the C-C$_0$,C-C$_{\pi}$ magnons with one gap minimum pinned at $ (0,0) $ or $ (0,\pi) $,
     or C-IC magnons with  two continuously changing gap minima  $ (0, \pm k^{0}_y ) $   tuned by the SOC strength
     can be measured by the corresponding peak positions of the uniform transverse Bragg spectroscopy at low temperatures \cite{lightatom1,lightatom2}.

     The interesting phenomena of one central peak $  S_{u,U(1)}^{+-}(\beta, \mathbf{k}) $ splits into two as tuning
     the gauge parameter $ \beta $ resembles those in the angle resolved photo emission spectrum (ARPS)
     $  S_{1}( \mathbf{k}, \omega ) $ as one tunes the momentum ( or energy ) in electron-hole semi-conductor bilayer \cite{ehbl1,ehbl2,ehbl3}
     or the  differential conductance $ \frac{ d I ( \mathbf{Q}, V ) }{ d V } $ as tuning the in-plane magnetic field
     $ \mathbf{Q}= \frac{ 2 \pi d }{\phi_{0} } B_{||} \hat{x} $ in the bilayer quantum Hall systems at total filling factor $ \nu_T= 1$ \cite{blqhrev}.
     In all the three systems, there is a pinned flat regime in the corresponding tuning parameters $ \beta, \mathbf{k}, \mathbf{Q} $
     before the single peak splits into two symmetric peaks with smaller heights.

{\sl 3.  Ground state and magnetization detection in longitudinal spin correlation functions and spin structure factors }

      We can also obtain the longitudinal spin-spin correlation functions.
     They are related to Eq.\eqref{lowzz} by the local spin rotation  $ \tilde{\mathbf{S}}_n = R( \hat{x}, \pi n_1 ) \mathbf{S}_n $
     which leads to a very simple relation between the two basis:
\begin{equation}
\begin{aligned}
      S_{u,U(1)}^{zz}(\mathbf{k},\omega) &= S_{s}^{zz}(\mathbf{k},\omega),\\
      S_{s,U(1)}^{zz}(\mathbf{k},\omega) &= S_{u}^{zz}(\mathbf{k},\omega)
\label{exch}
\end{aligned}
\end{equation}
     namely, there is an exchange between uniform and staggered components.
     Obviously, the uniform $ \chi_{u,U(1) } =  S_{u,U(1)}^{zz}(\mathbf{k} \rightarrow 0 ,\omega=0) $ and the staggered
     susceptibilities $ \chi_{s,U(1) }= S_{s,U(1)}^{zz}(\mathbf{k} \rightarrow 0 ,\omega=0) $
     will exchange their roles between the original and the $ U(1) $ basis.
     So they just show a broad distribution, in sharp contrast to the transverse dynamic correlation functions Eq.\eqref{lowpmu1}.

     Similarly, the equal-spin longitudinal structure factors  $ S_{u,U(1)}^{zz}(\mathbf{k})= S_{s}^{zz}(\mathbf{k}), S_{s,U(1)}^{zz}(\mathbf{k})= S_{u}^{zz}(\mathbf{k}) $ given in Eq.\eqref{lowzzequalleading} also display a broad distribution.

    Note that in contrast to the original basis discussed in Sec.V, now the magnetization part $ M^{2}(T) \delta_{ \mathbf{k},0} 2 \pi \delta( \omega ) $
    ($  M^{2}(T) \delta_{ \mathbf{k},0} $) due to the quantum  ground state in Fig.2a appear in the
    the uniform {\sl connected } dynamic (equal-time) longitudinal spin-spin correlation function $ S_{u}^{zz}(\mathbf{k},\omega) $
    ($ S_{u}^{zz}(\mathbf{k}) $) which can be detected easily by elastic longitudinal Bragg spectroscopy peak at momentum $ (0, 0) $
     in the RBZ at low temperatures \cite{lightatom1,lightatom2}.

\subsection{ Spin structure factors at high temperatures: high temperature expansions }

    At high temperature, even the magnetization vanishes, the Hamiltonian Eq.\eqref{u1h} in the U$(1)$ basis still break the lattice into two sublattices
    A and B shown in Fig.2a, so one still need to calculate the uniform and staggered spin structure factors separately.
    We get the uniform and staggered structure factors upto the order of  $ (J/T)^2 $:
\begin{eqnarray}
S_{u,U(1)}^{+-}(\mathbf{k})  =
	\left(\frac{J}{4T}-\frac{J^2}{16T^2}\right)[\cos k_x+\cos2\beta\cos k_y] ~~
              \nonumber  \\
	 +  \frac{J^2}{16T^2}[\cos2k_x+\cos4\beta\cos2k_y]   ~~~~~~~~~~~~~  \nonumber  \\
	 +  \frac{J^2}{8T^2}\cos2\beta[\cos(k_x+k_y)+\cos(k_x-k_y)]   \nonumber  \\
	S_{s,U(1)}^{+-}(\mathbf{k}) =
	\left(\frac{J}{4T}-\frac{J^2}{16T^2}\right)[-\cos k_x+\cos2\beta\cos k_y]
        \nonumber  \\
	+  \frac{J^2}{16T^2}[\cos2k_x+\cos4\beta\cos2k_y]  ~~~~~~~~~~~~~~ \nonumber  \\
	- \frac{J^2}{8T^2}\cos2\beta[\cos(k_x+k_y)+\cos(k_x-k_y)]  ~~~~~
\end{eqnarray}
     which are indeed different from Eq.\eqref{hightna} in the original basis.  Both depend on $ \beta $ explicitly and can be
     measured by Bragg spectroscopy experiments \cite{lightatom1,lightatom2}.

      We can also obtain the longitudinal spin structure factors which
      is related to Eq.\eqref{hightl} by the local spin rotation  $ \tilde{\mathbf{S}}_n = R( \hat{x}, \pi n_1 ) \mathbf{S}_n $.
      Then just similar to the low temperatures, we find again there is an exchange between uniform and staggered components
      in the two basis: $ S_{u,U(1)}^{zz}(\mathbf{k})= S_{s}^{zz}(\mathbf{k}), S_{s,U(1)}^{zz}(\mathbf{k})= S_{u}^{zz}(\mathbf{k}) $
      listed in Eq.\eqref{hightlsplit}.
      So they are also independent of the gauge parameter $ \beta $ up to the second order of $ (J/T)^2 $.

\subsection{ Comments on the finite temperature phase transitions }

     In Fig.3, there is a finite temperature transition from the $Y$-$x$ state to the paramagnet.
     However, because the $Y$-$x$ state is a spin-orbital correlated ground state which breaks
     both spin and translational symmetry. So in the original basis,
     it is not clear if the $Y$-$x$ to the paramagnet transition in Fig.3 will split into two transitions which restore the
     magnetization symmetry breaking and lattice symmetry breaking separately.
     However, this ambiguity can be resolved in the U$(1)$ basis. Because the $ Y $ ferromagnetic ground state only
     breaks the magnetization symmetry, so there can only be one transition to restore this symmetry breaking.
     The absolute value of the magnetization and  specific heat in Eq.\eqref{mc} are gauge invariant, they will display
     the critical behaviors $ C(T) \sim |T-T_c|^{-\alpha}, M(T) \sim |T-T_c|^{-\beta} $ with $ \alpha, \beta $ two critical exponents.
     The gauge invariance proves
     there can only be one transition in the original basis Fig.3.

     However, as emphasized in Sec.III, the Hamiltonian at $ \beta=\pi/4 $ has an extra symmetry which is broken
     by the $Y$-$x$ state. This extra symmetry breaking
     is important to determine the universality class of the C-IC to the paramagnet transition at $ \beta=\pi/4 $ in Fig.3.
     In fact, it controls the universality class of the whole phase boundary $ T_c(\beta ) $ in Fig.3.
     All the RG fixed points are shown in Fig.3: $ (\beta=\pi/4,T=T_m ) $ controls the finite temperature transition from
     the $Y$-$x$ state to the paramagnet state. $ ( \beta=\pi/2, T=0 ) $ controls the whole low temperature $Y$-$x$ phase.
     Of course, there is a fixed point at $ ( \beta=\pi/2, T=\infty ) $ controls the whole high temperature paramagnet phase.
     Determining the universality class of the finite temperature phase transition in Fig.3 remains an important outstanding problem.
     It could be related to the central charge $ c\leq 1 $ conformal field theory with the orbifold construction ( Note that Ising model is only $ c=1/2 $ ) \cite{kondo1,kondo2,cft,un1}.

\begin{figure}
\includegraphics[width=8.5cm]{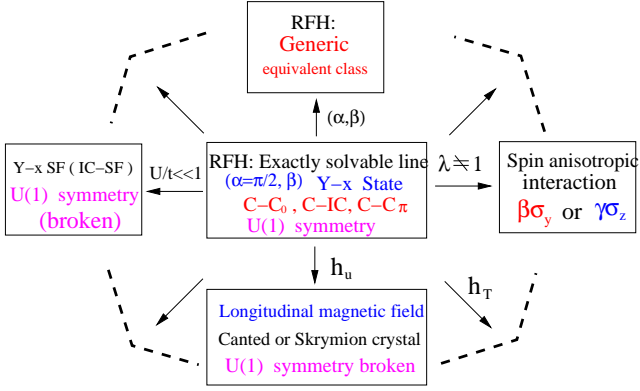}
\caption{ (Color online)
Adding or tuning various parameters
away from the solvable line $ ( \alpha=\pi/2, \beta ) $,
one can study various new quantum phases with different spin-orbital structures and quantum phase transitions among these phases.
Note that for $ \lambda \neq 1 $, there are still two different ways to put the non-abelian gauges fields: $ \beta \sigma_y $ to
break the U$(1)$ symmetry explicitly, another $ \gamma \sigma_z $ to
keep the U$(1)$ symmetry which maybe broken spontaneously by some canted or Skyrmion crystal states.
Adding a Zeeman field $ h_u $ or a transverse field $ h_T $ will also lead to quite different phenomena \cite{zeeman}. }
\label{global}
\end{figure}

\section{ Conclusions and perspectives on moving away from the solvable line}

 In this paper, we show that new class of quantum magnetism can be realized by strongly interacting spinor bosons loaded on optical lattices subject to non-Abelian gauge potentials. This new quantum magnetism can be captured by the Rotated Heisenberg model Eq.\eqref{rh} which may also be used to describe some materials with strong SOC or DM interaction.  Along the dashed line in Fig.1b, it displays a new class of commensurate spin-orbital correlated quantum phase
 with new elementary excitations (named as incommensurate magnons here) and phase transitions at finite temperatures.
 Although we achieved all these results to the leading order in the $ 1/S $ expansion, we expect all the results at $ T=0 $ are exact.
     Because the $ Y-x $ state
     is the exact eigenstate with no quantum fluctuations, there are no higher order corrections at $ T=0 $.
     so the excitation spectrum  of the C-C$_0$, C-IC, C-C$_{\pi}$ magnons in Fig.2 are exact.
     Their boundaries  $ \beta_1 $ and $ \beta_2 $ between the
      C-C$_0$, C-C$_{\pi}$ and  C-IC in Fig.2 are also exact.
     However, at small finite temperatures, there are higher order corrections due to the interactions among the magnons
     to all the physical quantities studied in this paper, which are expected to be small and can be evaluated straightforwardly.

 Our approach is from the three routes: (1) Exact statements from the symmetries, Wilson loops, gauge invariance and gauge transformations analysis (2)
 A well controlled SWE to leading order in $ 1/S $ at low temperature.
 (3) a well controlled high temperature expansion at high temperatures.
 Obviously, detailed calculations in (2) and (3) have to satisfy the constraints set by (1), which has been confirmed through the whole paper.
 Unfortunately,  both the low temperature SWE in (2) and the high temperature expansion in (3) fail near the finite temperature phase transition in Fig.3 whose
 universality class remains to be determined. Numerical calculations are needed to calculate all the physical quantities near the transition.

 It is instructive to compare with in-commensurability appeared in other lattice systems.
 In \cite{tqpt}, the authors investigated the topological quantum phase transition (TQPT)  of non-interacting
 fermions hopping on a honeycomb lattice in the presence of a synthetic non-Abelian gauge potential. The TQPT is driven by the collisions
 of two Dirac fermions located at in-commensurate momentum points continuously tuned by the  non-Abelian gauge parameters.
 The present paper focused on the strong coupling $ U/t \gg 1 $ limit along the solvable line.
 At weak coupling $ U/t $ limit along the solvable line, Eq.\eqref{bosonint} is expected to be in a superfluid (SF) state. In \cite{un1}, we will show that
 as one changes the gauge parameter $ \beta $ along the dashed line $ ( \alpha=\pi/2, \beta ) $ in Fig.1b, the system will undergo a
 C-IC transition from a C-SF state with $Y$-$x$ spin-orbital order to an IC-SF with in-commensurate spin-orbital orders which breaks
 both off-diagonal long range order and also the  U$(1)$ symmetry.
 The symmetry breaking lead to two gapless modes inside the IC-SF phase (Fig.10).

 It is also instructive to compare the C-IC magnons at $ ( 0, \pm k^{0}_y ) $ in Fig.2b in a lattice system with the roton minima in
 a continuous system. In the superfluid $ ^{4} He $ system, the roton inside the superfluid state indicates the short-ranged
 solid order embedded inside the off diagonal long-ranged SF order \cite{sf1,sf2,sf3}. As the pressure increases, the roton minimum drops and signals a
 first order transition to a solid order (or a putative supersolid order).  Similarly, the roton dropping in a 3d superconductor subject to a Zeeman field
 signals a transition from a normal state to the FFLO state \cite{fflo,fflolong,ssrev}.
 In 3d, the roton sphere is a 2d continuous manifold,
 so its dropping before touching zero signals a first order transition. Similarly, in a 2d electron-hole semi-conductor bilayer (EHBL) system \cite{ehbl}
 or 2d bilayer quantum Hall (BLQH) systems \cite{blqh1,blqh2,blqh3}, the roton circle is a 1d continuous manifold tuned by the distance between the two layers,
 so its dropping before touching zero also signals a first order transition.
 In contrast, the C-IC magnons in Fig.2b  are located at two isolated points $ ( 0, \pm k^{0}_y ) $,
 they indeed touch zero at all the transitions shown in Fig.10, so it signals a second order phase transition.

 The existence of the incommensurate magnons  above a commensurate phase is a salient feature of the RH model.
 They indicate the short-ranged in-commensurate order embedded in a long-range ordered commensurate ground state.
 Under the changes of the gauge parameters   $ (\alpha, \beta ) $, namely at generic equivalent classes (Fig.10),
 they are the seeds driving the transitions from commensurate to another commensurate phase with different spin-orbital
 structure or to an In-commensurate phase in the most general RH model Eq.\eqref{rh}.
 The effects of the spin-anisotropy interaction $ \lambda \neq 1 $ in Eq.\eqref{anisotropy} and the behaviors of the RFH
 in the presence  of external Zeeman fields will be discussed in separate publications \cite{un1,zeeman}.
 Preliminary results show that indeed the C-C$_0$, C-C$_{\pi}$ and  C-IC  magnons are the seeds to drive various
 quantum phase transitions under the effects of spin-anisotropy $ \lambda $ and the external magnetic fields $ \Omega $ (Fig.10).
 Especially, various different kinds of in-commensurate Skyrmion crystal phases  breaking the U$(1)$ symmetry, therefore leads to
 gapless Goldstone modes are identified.
 We expect that investigating the behaviors of this new elementary excitations in the RH model when tuned away from the solvable line  holds the key to
  explore all the possible fantastic new class of magnetic phenomena in materials with SOC or DM interaction.
  Rotated Anti-ferromagnetic Heisenberg model (RAFH) (not shown in Fig.10) which show dramatically different quantum phenomena will
  be presented elsewhere \cite{un1}.
  The RFH and RAFH models could be used to explore new class of magnetic phenomena in strongly correlated materials with
  strong SOC such as rare-earth insulators or iridium oxides.

{\bf Acknowledgements:}
We thank I. Bloch, Ruquan Wang and Jun Ye for helpful discussions on current and near future experimental status.
This research is supported by NSF-DMR-1161497, NSFC-11174210.
WL was supported by the NKBRSFC under grants Nos. 2011CB921502, 2012CB821305, NSFC under grants
Nos. 61227902, 61378017, 11434015, SPRPCAS under grants No. XDB01020300.

Note added: Very recently, a new experiment realizing a 2d Rashba SOC in $^{40}$K  Fermi gas came out \cite{expk40}.
Based on a simplified version of the proposal \cite{expliu}, a experiment to realize 2d Rashba SOC in a square optical lattice  is also undergoing \cite{private}.

\appendix

 In appendix A, we analyze the symmetries of the bosonic model Eq.\eqref{bosonint} and the RFH Eq.\eqref{rh} at the solvable line $ ( \alpha =\pi/2, \beta ) $,
 also the enlarged symmetry at $ \beta =\pi/4 $, the symmetry breaking patterns of the $Y$-$x$ state. In appendix B, we show that the
 $Y$-$x$ state is the exact ground state along the solvable line. In appendix C, we derive the exact constraints of the U$(1)$ symmetry on the spin correlation functions in both the original and U$(1)$ basis. In appendix D, establish the exact relations  between spin correlation functions in the original basis and those in the U$(1)$ basis due to the unitary transformation $ \tilde{b}_i=(i\sigma_x)^{i_x}b_i  $
 in the bosonic language or  $ \tilde{\mathbf{S}}_n = R( \hat{x}, \pi n_1 ) \mathbf{S}_n $  in the spin language.

\section{ Symmetry and symmetry breaking analysis }

    Along the dashed line $ ( \alpha =\pi/2, \beta ) $ in Fig.1b, the bosonic model Eq.\eqref{bosonint} has Time reversal $ \mathbf{k} \rightarrow -\mathbf{k}, \mathbf{S} \rightarrow -\mathbf{S} $, translational symmetry and three spin-orbital coupled $ Z_2 $ symmetries:
    (1) $ {\cal P}_x $ symmetry: $ S^{x} \rightarrow S^{x}, k_y \rightarrow - k_y, S^{y} \rightarrow - S^{y},  S^{z} \rightarrow - S^{z} $.
    (2) $ {\cal P}_y $ symmetry: $ S^{y} \rightarrow S^{y}, k_x \rightarrow - k_x, S^{x} \rightarrow - S^{x},  S^{z} \rightarrow - S^{z} $.
    (3) $ {\cal P}_z $ symmetry: $ k_x \rightarrow - k_x, S^{x} \rightarrow - S^{x}, k_y \rightarrow - k_y, S^{y} \rightarrow - S^{y},
    S^{z} \rightarrow S^{z} $ which is also equivalent to a joint $ \pi $ rotation of the spin and orbital  around $ \hat{z} $ axis.
    Most importantly, there is also a spin-orbital coupled U$(1)$ symmetry $[H_b,\sum_i(-1)^{i_x}b_i^\dagger\sigma^y b_i]=0 $.
    Of course, at the two Abelian points, the U$(1)$ symmetry is enlarged to the SU$(2)$ symmetry in the corresponding rotated basis.
    The $Y$-$x$ ground state in the Fig.2a breaks all these discrete symmetries except the $ {\cal P}_y $ and the U$(1)$ symmetry.
    It is two-fold degenerate.

    In the strong coupling limit, along the dashed line $ ( \alpha =\pi/2, \beta ) $, after rotating spin axis from $ Y $ to $ Z $, we reach
    the RH model Eq.\eqref{originalH}. It  has the Time reversal symmetry $ S^{z} \rightarrow - S^{z}, S^{+} \leftrightarrow -S^{-}, i \rightarrow -i $
    (here  $ i $ is the imaginary unit, not the site index).
    Translational symmetry and  the three spin-orbital coupled $ Z_2 $ symmetry:
    (1) $ {\cal P}_x $ symmetry:  $ S^{z}_j \rightarrow - S^{z}_{\bar{j}}, S^{+}_j \leftrightarrow S^{-}_{\bar{j}} $.
    where $ \bar{j} $ is the image of the site $ j $ reflected with respect to $ x $ axis.
    (2) $ {\cal P}_y $ symmetry:  $ S^{z}_j \rightarrow - S^{z}_{\bar{j}}, S^{+}_j \leftrightarrow - S^{-}_{\bar{j}} $.
    where $ \bar{j} $ is the image of the site $ j $ inverted with respect to the origin.
    (3) $ {\cal P}_z $ symmetry:  $ S^{z}_j \rightarrow S^{z}_{\bar{j}}, S^{+}_j \leftrightarrow -S^{+}_{\bar{j}} $.
    where $ \bar{j} $ is the image of the site $ j $ reflected with respect to $ y $ axis.
    Most importantly, there is also a spin-orbital coupled U$(1)$ symmetry $ [ H_{RH}, \sum_i(-1)^{i_x} S^{z}_{i} ]=0 $.
    Of course, at the two Abelian points, the symmetry is enlarged to SU$(2)$ symmetry in the corresponding rotated basis.
    The $Z$-$x$ ground state in the Fig.2a breaks all these discrete symmetries
    except the  $ {\cal P}_z $ and the  U$(1)$ symmetry. It is two-fold degenerate.

    It can be shown that under the local rotation $ \tilde{\mathbf{S}}_{i} =R(\hat{x},\pi ) R(\hat{y},\pi n_2) \mathbf{S}_{i}$, $ \beta \rightarrow \pi/2 - \beta $.
    The most frustrated point with $ W_R=-1 $ is  located at the middle point $ \beta=\pi/4 $ (Fig.1b) where the Hamiltonian has
    an extra symmetry invariant under $ \tilde{\mathbf{S}}_{i} = R(\hat{x},\pi ) R(\hat{y},\pi n_2) \mathbf{S}_{i} $.
    This extra symmetry is broken by the $Y$-$x$ state.
    As discussed in Fig.3 and Sec.VIII-C, this extra symmetry breaking
    is important to determine the universality class of the C-IC to the paramagnet transition at $ \beta=\pi/4 $.

    When performing  the unitary transformation  from the original basis to the U$(1)$ basis by the unitary matrix
    $  U=\prod_n e^{i \frac{\pi}{2} \sigma_x n_1 } $ listed above Eq.\eqref{kinetic1},
    all the symmetry operators transform accordingly $ P \rightarrow U P U^{-1} $. See appendix D below.

\section{ Proof of the $Y$-$x$ state as the exact ground state of the Hamiltonian Eq.\eqref{originalH} }

 Intuitively, we write the state $ Y-x =\mid S \rangle_A\otimes\mid -S \rangle_B$.

{\sl Lemma 1}: The $ Y-x $ state is an eigenstate of the Hamiltonian.

Since site $i$ and $i+x$ belong to different sublattice,
without loss of generality we set $i\in A$,
then we have
\begin{eqnarray}
	S_i^+S_{i+x}^+|Y-x\rangle
	=0
    \nonumber   \\
	S_i^-S_{i+x}^-|Y-x\rangle
	=0
\end{eqnarray}
While site $i$ and $i+y$ belong to the same sublattice, if $  i \in A $, we have
\begin{eqnarray}
	S_i^+S_{i+y}^-|Y-x\rangle
	=0
 \nonumber   \\
	S_i^-S_{i+y}^+|Y-x\rangle
	=0
\end{eqnarray}
  Same calculations hold for $i\in B$. In all
\begin{equation}
	H|Y-x\rangle = -2NJS^2|Y-x\rangle
\end{equation}

{\sl Lemma 2}: The $ Y-x $ state saturates the lower bound of the Hamiltonian.

For a given bond from $i$ to $j$, since $R\in SO(3)$,  one can introduce $\tilde{S}_j^a=R^{ab} S_j^b$,
then
\begin{equation}
	-S(S+1)\leq \langle S_i^a R^{ab} S_j^b\rangle\leq S^2 ~~~~~~~~~
\end{equation}
 which leads to the lower bond:
\begin{equation}
	\min\langle H\rangle \geq  -2NJS^2
\end{equation}

  Combining Lemma 1 and Lemma 2 concludes that the state $Y-x $ is indeed the ground state.
  Obviously, the ground state has two-fold degeneracy which are related by the Time Reversal, or translation by one lattice site,
  or by the  spin-orbital coupled $ Z_2 $ symmetries $ {\cal P}_x $ or $ {\cal P}_y $ of the Hamiltonian.

    As shown in Sec.IV, the  $ Y-x $ ground state and the magnetization $ M (T) $ Eq.\eqref{mc} can be determined by the
    sharp peak and its spectral weight of Bragg spectroscopy
    in the staggered {\sl longitudinal} spin-spin correlation function at low temperatures.

\section{ The exact constraints of the U$(1)$ symmetry on the spin correlation functions in the original and  U$(1)$ basis. }

{\sl 1. The original basis: }

  The U$(1)$ symmetry operator in the original basis is $ U_{1}(\alpha) = e^{ i \alpha \sum_{i} (-1)^x S^{z}_{i} } $.
  It is easy to see that
\begin{eqnarray}
	U_{1}(\alpha) S^{\pm}_{A}(\mathbf{k}) U^{-1}_{1}(\alpha) =e^{\pm i \alpha }  S^{\pm}_{A}(\mathbf{k})
 \nonumber   \\
	U_{1}(\alpha) S^{\pm}_{B}(\mathbf{k}) U^{-1}_{1}(\alpha) =e^{\mp i \alpha }  S^{\pm}_{B}(\mathbf{k})
\end{eqnarray}
    Using the definition $ S^{\pm}_u= S^{\pm}_{A} + S^{\pm}_{B} $, the ground state $ | G \rangle $ is the U$(1)$ invariant   $
    | G \rangle = U_{1}(\alpha) | G \rangle  $ and $ [U_1(\alpha), H]=0 $, one can show that the invariance of $ S^{+-}_{u}( \mathbf{k}, t ) $
    under the U$(1)$ symmetry dictates:
\begin{equation}
	\langle  S^{+}_{A}(\mathbf{k},t) S^{-}_{B}(\mathbf{k}, 0) \rangle= \langle  S^{+}_{B}(\mathbf{k}, t) S^{-}_{A}(\mathbf{k}, 0) \rangle =0
\end{equation}
    which leads to
\begin{eqnarray}
	S^{+-}_{u}( \mathbf{k}, t ) & = & \langle  S^{+}_{A}(\mathbf{k}, t) S^{-}_{A}(\mathbf{k}, 0) \rangle + \langle  S^{+}_{B}(\mathbf{k},t) S^{-}_{B}(\mathbf{k}, 0) \rangle
                            \nonumber  \\
      & = &   S^{+-}_{s}( \mathbf{k}, t )
\end{eqnarray}
    which justifies the first equation in Eq.\eqref{usus}.

    Similarly, the invariance of $ S^{++}_{u}( \mathbf{k}, t ) $
    under the U$(1)$ symmetry dictates that
\begin{equation}
	\langle  S^{+}_{A}(\mathbf{k},t) S^{+}_{A}(\mathbf{k}, 0) \rangle= \langle  S^{+}_{B}(\mathbf{k}, t) S^{+}_{B}(\mathbf{k}, 0) \rangle =0
\label{aa}
\end{equation}
    which leads to
\begin{eqnarray}
	S^{++}_{u}( \mathbf{k}, t ) & = & \langle  S^{+}_{A}(\mathbf{k}, t) S^{+}_{B}(\mathbf{k}, 0) \rangle + \langle  S^{+}_{B}(\mathbf{k},t) S^{+}_{A}(\mathbf{k}, 0) \rangle
           \nonumber  \\
      &= &  -S^{++}_{s}( \mathbf{k}, t )
\label{ab}
\end{eqnarray}
    which justifies the second equation in Eq.\eqref{usus}.

    The U$(1)$ symmetry also dictates that the correlation functions between the longitudinal spin and transverse ones vanish.

{\sl 2. The U$(1)$ basis: }

    Obviously, the U$(1)$ symmetry operator in the U$(1)$ basis is
    $ U U_{1}(\alpha) U^{-1}= e^{ i \alpha \sum_{i} S^{z}_{i} }= \tilde{U}_{1}(\alpha)$
    where $  U=\prod_n e^{i \frac{\pi}{2} \sigma_x n_1 } $ is the unitary transformation between the original basis and the  U$(1)$ basis
    listed above Eq.\eqref{kinetic1}.
    The $ \tilde{U}_{1}(\alpha) $ symmetry directly leads to Eq.\eqref{u1zero}.

\section{ The relations  between spin correlation functions in the original basis and those in the U$(1)$ basis. }

    Now we establish the connections between the correlation functions in the original basis and those in the U$(1)$ basis.
    In the original basis, $ H_d(S) |G\rangle  = E |G\rangle $ where the Hamiltonian $ H_d(S) $ is given by Eq.\eqref{originalH}.
    In the U$(1)$ basis, $ H_{U(1)}(S) |\tilde{G} \rangle= E |\tilde{G} \rangle $, the
    $ |\tilde{G} \rangle= U |G\rangle $ is  the ground state in the U$(1)$ basis and satisfies
    $ |\tilde{G} \rangle=\tilde{U}_{1}(\alpha) |\tilde{G} \rangle $,
    the $  H_{U(1)}(S)= U H_d(S) U^{-1}= H_d( U S U^{-1} )= H_d ( R(\hat{x},n_1 \pi) S ) $
    is given in Eq.\eqref{u1h}.

    Using the definition  $ \tilde{S}^{\pm}_{A,B}(t)= e^{i H_{U(1)}(S) t} S^{\pm}_{A,B}(t)  e^{-i H_{U(1)}(S) t} $,
    one can see that
\begin{eqnarray}
	U^{-1} \tilde{S}^{\pm}_{A}(\mathbf{k},t) U  = S^{\pm}_{A}(\mathbf{k},t )
 \nonumber   \\
	U^{-1} \tilde{S}^{\pm}_{B}(\mathbf{k},t) U  = S^{\mp}_{B}(\mathbf{k},t )
\end{eqnarray}

    After transferring from the U$(1)$ basis back to the original basis, one can show that
    Eq.\eqref{u1zero} leads to the constraints in the original basis:
\begin{eqnarray}
\langle  S^{+}_{A}(\mathbf{k},t) S^{+}_{A}(\mathbf{k}, 0) \rangle + \langle  S^{-}_{B}(\mathbf{k}, t) S^{-}_{B}(\mathbf{k}, 0) \rangle  =0
                          \nonumber   \\
\langle  S^{+}_{A}(\mathbf{k},t) S^{-}_{B}(\mathbf{k}, 0) \rangle + \langle  S^{+}_{B}(\mathbf{k}, t) S^{-}_{A}(\mathbf{k}, 0) \rangle =0
\end{eqnarray}
      which are consistent with the Eq.\eqref{aa},\eqref{ab} achieved in the original basis directly.

     Similarly,  after transferring from the U$(1)$ basis back to the original basis,
     one can show that Eq.\eqref{lowpmu1} leads to:
\begin{eqnarray}
     S^{+-}_{u/s,U(1)}( \mathbf{k}, t )=\langle  S^{+}_{A}(\mathbf{k},t) S^{-}_{A}(\mathbf{k}, 0) \rangle
      + \langle  S^{-}_{B}(\mathbf{k}, t) S^{+}_{B}(\mathbf{k}, 0) \rangle
                          \nonumber   \\
       \pm \langle  S^{+}_{A}(\mathbf{k},t) S^{+}_{B}(\mathbf{k}, 0) \rangle \pm \langle  S^{-}_{B}(\mathbf{k}, t) S^{-}_{A}(\mathbf{k}, 0) \rangle ~~~~~~~~~
\end{eqnarray}
     which, as stressed below Eq.\eqref{lowpmu1}, are not directly related to the corresponding transverse spin correlation functions
     Eq.\eqref{lowna} in the original basis.

     However, the longitudinal correlations functions in the two basis are simply related by
     Eq.\eqref{exch}.



\begin{thebibliography}{99}
\bibitem{direct}
More directly, one can just put $ \beta \sigma_z $  along the $ y $ bonds in Fig.1a.
If one still put $ \beta \sigma_y $  along the $ y $ bonds in Fig.1a, then the longitudinal spin-spin correlation function $ S^{zz}_{u,s} $
should be understood as $  S^{yy}_{u,s}  $.
\bibitem{drop}
For simplicity, we drop the $ \tilde \tilde $ on top of $ \mathbf{S}_n $.
\bibitem{note}
Compared to the HP bosons introduced in the original basis in the Sec.IV, one can see
the reverse of $ S^{z} \rightarrow - S^{z}, S^{+} \leftrightarrow S^{-} $ in the sublattice B.
\end{thebibliography}
\end{document}